\documentclass[aps, pra,
reprint,
superscriptaddress,
twocolumn,
floatfix]{revtex4-2}
\usepackage[T1]{fontenc}
\usepackage{float}
\usepackage[toc,page]{appendix}
\usepackage{amsmath,mathtools,amsthm,amssymb}
\usepackage{bm}
\usepackage{adjustbox}
 
\usepackage{tabularx}
\usepackage{units}
\usepackage{complexity}
\usepackage{graphicx}
\usepackage{color}
\usepackage{xcolor}
\usepackage{bbold}
\usepackage{complexity}
\usepackage{booktabs}
 
\usepackage[linesnumbered,ruled,vlined]{algorithm2e}
\usepackage{listings}

\usepackage{lipsum} 
\usepackage{enumitem}

\usepackage{pifont}
\usepackage{times}
\usepackage{comment}
\usepackage{array}
\newcolumntype{H}{>{\setbox0=\hbox\bgroup}c<{\egroup}@{}}
\usepackage{graphics}
\usepackage[normalem]{ulem}
\usepackage{complexity}

\definecolor{myrefcolor}{RGB}{242, 10, 10}
\definecolor{myurlcolor}{RGB}{255, 138, 48}

\usepackage[
    breaklinks,
    pdftex,
    colorlinks=true,
    linkcolor=myrefcolor,
    citecolor=myurlcolor,
    urlcolor=myurlcolor
]{hyperref}

\usepackage[toc,page]{appendix}

\usepackage{braket}
\usepackage{physics}
\usepackage{csquotes}
\usepackage{thmtools, thm-restate}

\definecolor{antonio}{rgb}{.2,.5,.1}

\usepackage{float}
\usepackage{qcircuit}

\newcommand{\dc}{\Delta_{\text{\scriptsize cp}}} 
\newcommand{\da}{\Delta_{\text{\scriptsize ap}}}

\usepackage{array}
\newcolumntype{P}[1]{>{\raggedright\let\newline\\\arraybackslash\hspace{0pt}}p{#1}}

\newcommand{\Bigb}[1]{\Big( #1\Big)}

\newcommand{\Bigbl}[1]{\Big[ #1\Big]}
\newcommand{\kx}{\kappa_{\text{\scriptsize ex}}}
 \usepackage{xcolor}
 
\usepackage{longtable}

\usepackage{url}

\begin{document}
\preprint{ }
\title{A vapor-cavity-QED system for quantum computation and communication}
\author{Sharoon Austin}
  \affiliation{Joint Quantum Institute, NIST/University of Maryland, College Park, MD, 20742, USA}\affiliation{Joint Center for Quantum Information and Computer Science, NIST/University of Maryland, College Park, MD, 20742, USA}
\author{Dhruv Devulapalli}  
\affiliation{Joint Quantum Institute, NIST/University of Maryland, College Park, MD, 20742, USA}\affiliation{Joint Center for Quantum Information and Computer Science, NIST/University of Maryland, College Park, MD, 20742, USA}
\author{Khoi Hoang}
  \affiliation{Joint Quantum Institute, NIST/University of Maryland, College Park, MD, 20742, USA}
\affiliation{Microsystems and Nanotechnology Division, Physical Measurement Laboratory, National Institute of Standards and Technology,
Gaithersburg, Maryland 20899, USA}
\author{Feng Zhou}
\affiliation{Joint Quantum Institute, NIST/University of Maryland, College Park, MD, 20742, USA}
\affiliation{Microsystems and Nanotechnology Division, Physical Measurement Laboratory, National Institute of Standards and Technology,
Gaithersburg, Maryland 20899, USA}
\author{Kartik Srinivasan}
\affiliation{Joint Quantum Institute, NIST/University of Maryland, College Park, MD, 20742, USA}
\affiliation{Microsystems and Nanotechnology Division, Physical Measurement Laboratory, National Institute of Standards and Technology,
Gaithersburg, Maryland 20899, USA}
\author{Alexey V. Gorshkov}
\affiliation{Joint Quantum Institute, NIST/University of Maryland, College Park, MD, 20742, USA}\affiliation{Joint Center for Quantum Information and Computer Science, NIST/University of Maryland, College Park, MD, 20742, USA}

\begin{abstract}
In this work, we propose performing key operations in quantum computation and communication using room-temperature atoms moving across a grid of high-quality-factor, small-mode-volume cavities. These cavities enable high-cooperativity interactions with single atoms to be achieved with a characteristic timescale much shorter than the atomic transit time, allowing multiple coherent operations to take place. We study scenarios where we can drive a Raman transition to generate photons with specific temporal shapes and to absorb, and hence detect, single photons. The strong atom-cavity interaction can also be used to implement the atom-photon controlled-phase gate, which can then be used to construct photon-photon gates, create photonic cluster states, and perform non-demolition detection of single photons. We provide numerics validating our methods and discuss the implications of our results for several applications. 
\end{abstract}

 \volumeyear{ }
 \volumenumber{ }
\issuenumber{ }
\eid{ }
\date{\today}
\startpage{1}
\endpage{10}
\maketitle

\section{Introduction}

The ability to create single photons, implement photon-photon gates, create photonic cluster states, and detect single photons is important for performing quantum computation \cite{pan_multiphoton_2012, kok_linear_2007}, quantum communication \cite{qkdlk}, and quantum metrology \cite{friis_flexible_2017}. Moreover, it is important to execute these operations in a scalable way to produce large-scale systems capable of implementing quantum error correction and solving large problems \cite{knill_resilient_1998}.

Single-photon sources have been realized using quantum dots \cite{qdlow} and atoms \cite{axelnet} in high-finesse cavities. Moreover, 
the use of auxiliary classical fields to control the shape \cite{waveform, qdwave} and polarization \cite{polalt} of the emitted photons has also been demonstrated.
However, spectral inhomogeneities in quantum dots \cite{inhom} and the technical demands to cool and trap atoms \cite{trp1, trp2} make scalability challenging. Similarly, state-of-the-art single-photon detectors based on superconducting nanowires require cryogenic operating temperatures \cite{snspd}.

\begin{figure*}[t!]
	\begin{center}
            \includegraphics[
		width=2 \columnwidth
		]{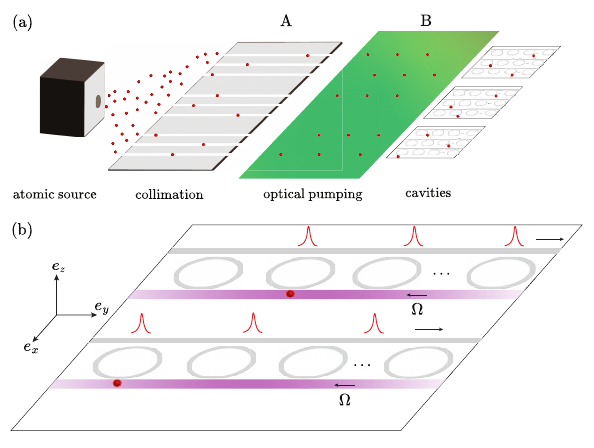}
	\end{center}
    	\caption{A schematic of the envisioned vapor cavity QED architecture. (a) The atoms produced from a thermal source go through two stages before entering the microcavity chips. In stage A, the atoms pass through a beam collimator to limit their transverse velocity. In stage B, the atoms are optically pumped using a laser (shown in green) so that they are initialized in the correct atomic state. (b) A schematic of the cavity QED chip. The atoms, moving in a collimated beam parallel to the waveguides, couple to microcavities arranged in a grid. This figure shows the generation of single photons (red wavepackets) with the help of free-space control pulses $\Omega$ (shown in purple). All other operations discussed in this work can be implemented similarly. Moreover, while microring whispering-gallery mode cavities \cite{md1, ringrev} are depicted, our approach applies to other geometries, such as racetrack microcavities \cite{race} and single-sided photonic-crystal cavities \cite{tiecke_nanophotonic_2014}.}
	\label{arch}
\end{figure*}

In this paper, we propose an architecture, shown schematically in Fig.~{\ref{arch}}, 
that can realize key photon processing primitives---photon sources, photon detectors, and photon-photon gates---without the physical infrastructure associated with laser cooling and trapping or cryogenics, and with a level of homogeneity that exceeds that of most solid-state quantum optical systems (such as quantum dots). Our architecture uses room-temperature atoms moving across a grid of high-quality-factor, small-volume cavities in a regime where strong atom-cavity interactions (large single-atom cooperativity) can be achieved. Previous work with cold atoms in cavity QED (quantum electrodynamics) systems has demonstrated all the key steps in implementing the prerequisite strong atom-light interactions \cite{boca_observation_2004, MOT_dayan, nonlinearvolz,tiecke_nanophotonic_2014,Reiserer2014,samutpraphoot_strong_2020}. In contrast, warm atoms present an additional challenge due to the finite time over which they are present within the cavity mode volume. Thus, to make use of this architecture for photonic quantum information processing, it is important that the atom's transit time through a cavity is much longer than the time it takes to generate a single photon, detect a single photon, or implement an atom-photon quantum gate. As a result, while cavity QED with warm atoms was pursued in groundbreaking experiments decades ago~\cite{thompson_observation_1992,turchette_measurement_1995}, the aforementioned challenge pointed towards cold-atom cavity QED as a more fruitful path for realizing coherent single-photon operations.

In this paper, we argue that developments in integrated photonics and microfabricated atomic devices suggest that cavity QED with warm vapors is worth revisiting, as recently noted in Ref.~\cite{alaien}. In particular, there has been continued development of integrated photonic microcavities with ultra-small mode volumes and high quality factors~\cite{pht1,alaien,chang_efficiently_2020,lu_high}, so that high-cooperativity atom-photon interactions are achievable. Moreover, recent work 
on microfabricated atomic beam collimators~\cite{colim} has shown that they can be incorporated together with microfabricated atomic vapor sources~\cite{martinez_chip-scale_2023}, thereby limiting the transverse velocity of the atoms. Taking these developments together, the timescale for coherent atom-photon interactions can be more than two orders of magnitude shorter than the atomic transit time, so that an appreciable number of single-photon operations can occur while the atom is interacting with the cavity. From a technology perspective, in recent years, there have been several efforts to create hybrid platforms enabling near-field interactions between integrated photonics and atomic vapors in fully integrated systems~\cite{stern,ritter}, including recent experimental work demonstrating atom-cavity interactions at the level of a few atoms and a few intracavity photons~\cite{zektzer_strong_2024}. Given the ability to create large arrays of devices through nanofabrication, and the ability to operate warm atoms without the auxiliary infrastructure associated with cooling and trapping, we anticipate our approach will be particularly well-positioned to benefit from the use of multiplexing. 

In the sections that follow, we outline well-known vapor-cavity-QED-based protocols and optimize them for the parameter regimes suitable for our system. In Sec.~\ref{sec:vqed}, we describe in detail our setup based on vapor cavity QED, and in Sec.~\ref{sec:warm_atoms}, we summarize the considerations specific to working with warm atoms. In Sec.~\ref{singlephoton}, we discuss schemes in two different parameter regimes to generate and detect single photons. In Sec.~\ref{cpfsec}, we study the physics of the atom-photon gate. In Sec.~\ref{apps}, we discuss applications of the basic primitives, including making cluster states \cite{Lind, pich} and performing quantum communication protocols \cite{qkdlk}.
Our work complements other recent studies of warm-atom cavity QED systems, such as Ref.~\cite{alaien}, where the potential for observing vacuum Rabi splitting and single-atom transits through an ultra-small volume photonic crystal cavity was studied.

\section{Vapor Cavity QED System}\label{sec:vqed} 
The system we consider consists of a grid of chip-scale microcavities that are linked to a warm-atom source through an atomic beam collimator, as depicted in Fig.~\ref{arch}, 
and it is expected that the entire system can be micro/nanofabricated and linked together in a compact and deployable package, for example, as has recently been demonstrated for atomic beam clocks~\cite{martinez_chip-scale_2023}. 

\begin{figure}[t]
		\includegraphics[
		width=0.8\columnwidth
		]{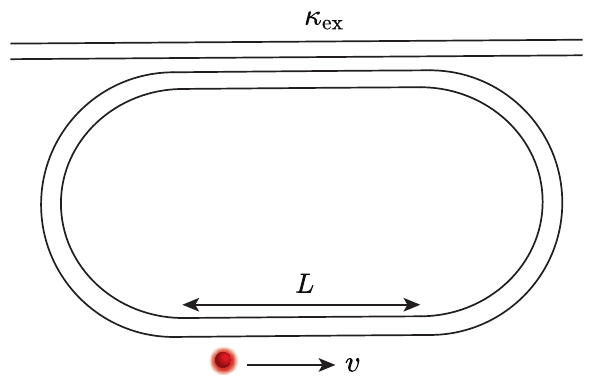}
	\caption{A racetrack microcavity, where an atom traverses the longest dimension with speed $v$, and the transverse velocities have been made sufficiently low through a prior stage of atomic beam collimation. Similar considerations for a single-sided Fabry-Perot cavity (e.g., based on a photonic-crystal geometry \cite{alaien}) hold.}
	\label{Racetrack}
\end{figure}

The cavities have ultra-low mode volumes and high quality factors that allow for strong atom-photon coupling, and the key parameters of the system include the single-photon Rabi frequency $g$, the rate $\kappa_{\text{ex}}$ at which the microcavity couples to the waveguide, the intrinsic loss $\kappa_{\text{i}}$ of the microcavity, and the decay rate $\gamma$ of the atomic coherence. Here $\tau$ is defined as the time during which the atom interacts with the cavity mode. Taking the racetrack microcavity as an example, $\tau$ is defined as the time it takes for the atom in a collimated beam to traverse the longest dimension, as shown in Fig.~{\ref{Racetrack}}.

Microcavities come in a variety of geometries, including microdisks \cite{md1}, microrings \cite{ringrev}, microbottles \cite{mb1}, and photonic-crystal defect structures \cite{alaien,thompson_coupling_2013, lu_high}. These microcavities couple to the waveguide and the atom through their evanescent modes. Moreover, the tight confinement of single-photon fields allows for ultra-small mode volumes and hence large single-photon electric fields and strong atom-field couplings \cite{kiprep, kiprep2}. 

\begin{figure}[t]
		\includegraphics[
		width=0.8\columnwidth
		]{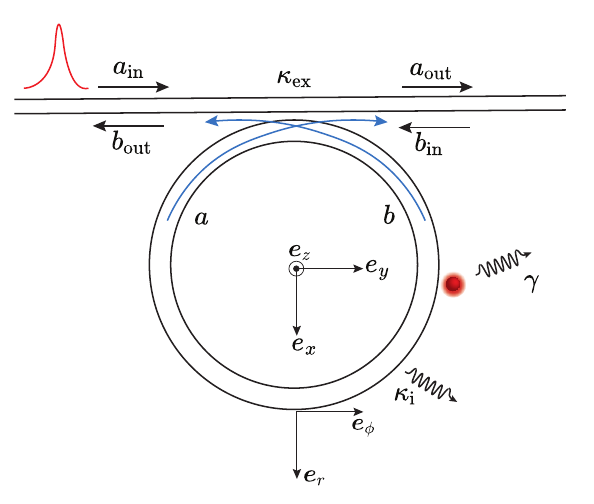}
	\caption{A schematic diagram of a microring, microbottle, or microtoroid cavity. Input fields $a_{\text{in}}$ and $b_{\text{in}}$ couple to the cavity with rate $\kappa_{\text{ex}}$. The cavity has an intrinsic loss rate $\kappa_{\text{i}}$. The input and output single-photon fields obey $a_{\text{out}}=a_{\text{in}}+\sqrt{2\kappa_{\text{ex}}}a$ and $b_{\text{out}}=b_{\text{in}}+\sqrt{2\kappa_{\text{ex}}}b$. In the perfectly  
    chiral case, the atom is coupled to
    the mode $a$ ($b$)
    which enables the $\sigma^{+}$ ($\sigma^-$) transition, and the atomic state is initialized to the lower level of the appropriate transition (see Ref.~\cite{Rauss} as an example). In the perfectly non-chiral case, the atom is coupled to both cavity modes which enable $\pi$ transitions between the two atomic states, and the atomic state is initialized to the lower level of the appropriate atomic transition.}
	\label{mt1}
\end{figure}

We consider two different polarizations of the modes of resonators as shown in Fig.~\ref{mt1}. For the perfectly chiral case, the optical modes $a$ and $b$ correspond to polarization $\bm{e}_{\sigma^{\pm}}$, enabling $\sigma^{\pm}$ transitions between two atomic levels. If we only excite the $a$ mode through $a_{\text{in}}$, as shown in Fig.~\ref{mt1}, we can ignore the dynamics of the $b$ mode (assuming $a$ and $b$ are not coupled). Moreover, the phase of the cavity electric field $\phi$, appearing in the atom-cavity coupling $g=\abs{g}e^{i\phi}$, can be absorbed into a redefinition of the cavity mode operator $a$. For the perfectly non-chiral case, the optical modes $a$ and $b$ correspond to polarization $\bm{e}_z$, enabling a $\pi$ transition between two atomic levels. In that case, as we will discuss below Eq.~(\ref{nonchiralH}) in Sec.~\ref{singlephoton}, the phase of the cavity field can also be absorbed into a redefinition of the optical modes, giving us a real $g$. 

We next qualitatively consider the extent to which real devices will exhibit perfect chirality/non-chirality. For photonic integrated circuit resonators such as those studied in this work, the transverse electric (TE) polarization is typically defined as having its dominant electric field component along the radial direction of the resonator ($\mathbf{e}_r$ direction in Fig.~\ref{mt1}), while the transverse magnetic (TM) polarization has its dominant electric field component orthogonal to the plane of the resonator (i.e., along $\mathbf{e}_z$ direction in Fig.~\ref{mt1}). Despite the above general designations, it has been shown that, depending on resonator geometry and refractive index, significant longitudinal electric field components can be present, enabling chiral behavior to be realized. Outside of photonic integrated circuits, this has been shown by Rauschenbeutel and colleagues for geometries such as microbottle resonators, where one of the two polarizations can have a significant longitudinal component, corresponding (approximately) to the perfectly chiral modes $a$, $b$ with polarizations $\bm{e}_{\sigma^{\pm}}$, while the other polarization is best described as the perfectly non-chiral modes $a$, $b$ with polarization $\bm{e}_z$ ~\cite{TM,TM2}. In contrast, for the higher-refractive-index-contrast, smaller-mode-volume systems that are the focus of our study, both polarizations can have significant longitudinal components due to strong transverse field confinement~\cite{espinosa-soria_transverse_2016,zektzer_chiral_2019,zhou_coupling_2023}. Therefore, the cavities considered here can have significant longitudinal field components with a $\pm \pi/2$ phase shift, so that, in general, the field is elliptically polarized, with specific spatial locations 
providing perfect circular polarization. For example, for the microring and microdisk resonators, due to the azimuthal symmetry, the degree of circular polarization varies in the $r$--$z$ plane.
Because our work is focused on thermal atomic beams, individual atoms within the beam will experience differing levels of circular polarization, so our considerations for the perfectly chiral and perfectly non-chiral configurations represent limiting cases. That being said, recent work has shown that through geometric engineering (e.g., of the waveguide cross-section), a strong overall degree of circular polarization within the evanescent tail of a propagating waveguide mode can be achieved, even in an average sense, with chiral effects in atomic vapors coupled to such waveguides shown~\cite{zektzer_chiral_2019}. From this point forward, we will use `chiral' to refer to the perfectly chiral case and `non-chiral' to refer to the perfectly non-chiral case.  

Designing an appropriate system for our applications will involve optimizing several design features. Using atomic beam collimators \cite{colim} and adjusting the cavity orientation accordingly ensures that the atoms traverse the longest dimension of the cavity. Choosing a specific geometry that maximizes the interaction length $L$ and hence the transit time $\tau \sim L$---such as in racetrack microcavities \cite{race} shown in Fig.~\ref{Racetrack} and 1D Fabry-Perot cavities with Bragg mirrors \cite{bragg}---comes at the cost of a lower $g$ because  $g \sim L^{-1/2}$. 
 As emphasized in the later sections, while the fidelity of our operations increases for larger values of $g$, we will have to balance this with a competing design goal of having a large enough $\tau$ so that a sufficient number of useful operations can be carried out per transit event.

\section{Considerations specific to warm atoms}\label{sec:warm_atoms}
In this section, we briefly summarize the main challenges in using warm atoms to implement the high-fidelity operations mentioned earlier. The first challenge is that warm atoms have a limited transit time, thereby limiting the number of useful atom-light interactions that can occur during a given atom's transit. This can be addressed by reducing the timescales over which our atom-light interactions take place, i.e.,~we choose values of $T$ (the pulse length of single photons) that are much smaller than the transit time $\tau$. For example, producing single photons at cavity resonance with high efficiency $\eta$ requires the adiabatic limit $\alpha \equiv \kappa T \gg 1$ (assuming $\kappa_i =0$ for simplicity, so $\kappa = \kappa_\text{ex}$), giving us (see Sec.~\ref{singlephoton})
\begin{align}
g^2 = \alpha \left(\frac{\gamma}{T}\right) \left( \frac{\eta}{1-\eta} \right) \gg \left(\frac{\gamma}{T}\right) \left( \frac{\eta}{1-\eta} \right).
\end{align}
This relation follows from expressing $g^2$ in terms of the cooperativity as $g^2=C\kappa \gamma$. Substituting $C=\eta/(1-\eta)$ (which follows from $\eta = C/(1+C)$), as discussed in Sec.~\ref{singlephoton}, and using $\alpha = \kappa T$ to eliminate $\kappa$ gives us the desired expression. 
Choosing both $T \ll \tau$ for room-temperature atoms and values of $\eta$ approaching 1 requires large values of $g$, which have only recently become routinely available \cite{largecoup}.
The second challenge is the temporal variation in coupling strength $g$ as the atom passes through the cavity mode. This is mitigated by using a tightly collimated beam \cite{colim} transiting over a particular orientation of the cavity, so the change in $g$ is minimal. The third challenge is that significant Doppler shifts of the cavity field and control pulses can decrease the fidelity of the proposed operations. We show in Sec.~\ref{expgf} how to ameliorate this issue.

\begin{figure}[t]
		\includegraphics[
		width= \columnwidth
		]{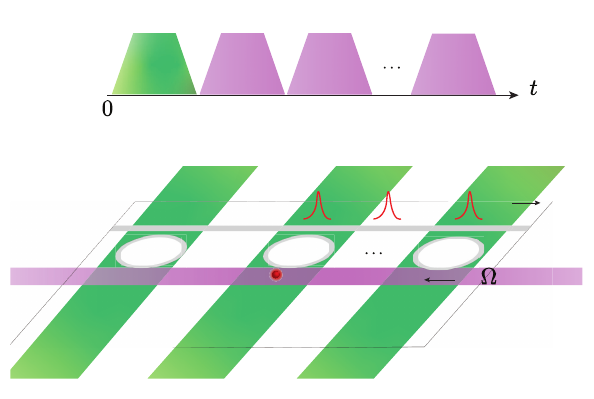}
	\caption{Before implementing any atom-photon interactions, classical laser pulses (shown in green) can be used to check which cavity is active and measure the single-photon Rabi frequency $g$. In this figure, we show the example of single-photon generation where the middle cavity is found to be active, and the remaining time slots are used for single-photon retrieval using the classical control pulses  $\Omega$.} 
	\label{checking}
\end{figure}

Since atom-cavity interactions will be non-deterministic, we propose using multiplexing to implement high-efficiency operations. Multiplexing schemes have already been developed to deal with the dead time of photodetectors.
For example, in spatial multiplexing \cite{meyer-scott_single-photon_2020}, a switching network monitors which detectors in a detector array have recently fired and routes the input pulse into those that have not. 
As shown in Fig.~\ref{checking}, the analogous procedure here would be to use classical pulses to detect which cavities are active (have an atom in them) and to route the incoming single photon to an active cavity. For single-photon sources, we can consider a protocol that detects the presence of an atom in the cavity prior to applying the control pulse, producing a heralded single photon. This is feasible since state-of-the-art on-chip modulators feature switching times of $<30$~ps \cite{hiswitch}, which is much shorter than the achievable transit time $\tau$ ($\sim 10$ ns, as shown in Table \ref{tab1})
and single-photon duration $T$ (few nanoseconds, as shown in Table \ref{tab1}). 
In addition, we can also consider checking whether a given cavity is still active at the end of our operation. We will also consider `passive' multiplexing wherein we can simply use a large number of cavities to make atom-photon interactions more probable, at the cost of effectively increasing the intrinsic loss $\kappa_{\text{i}}$. 
\section{Single-photon generation and detection}\label{singlephoton}

In this section, we outline the coherent control technique that can be used to generate and absorb single photons using a $\Lambda$-type atom coupled to a cavity \cite{made}. We also study in detail how various parameter regimes determine the properties of single photons that can be generated and absorbed. 

\begin{figure}[b]
		\includegraphics[
		width=0.6\columnwidth
		]{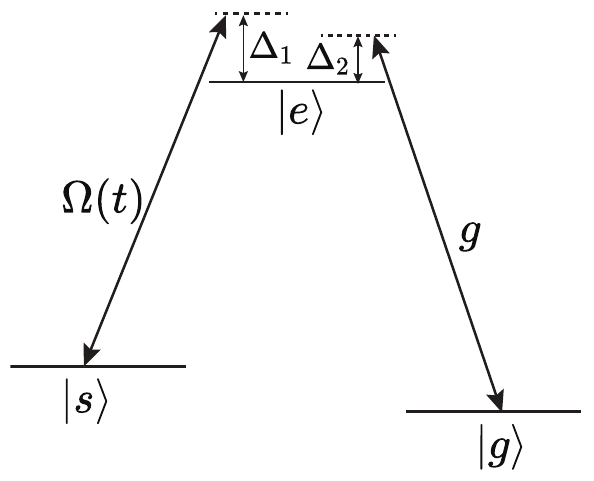}
	\caption{Atomic levels $\ket{s}$ and $\ket{e}$ are coupled by a classical laser pulse with Rabi frequency $\Omega(t)$, while atomic levels $\ket{e}$ and $\ket{g}$ are coupled by the cavity with single-photon Rabi frequency $g$. }
	\label{atomiclevel}
\end{figure}

As shown in Fig.~\ref{atomiclevel}, the cavity couples atomic states $\ket{e}$ and $ {\ket{g}}$ with single-photon Rabi frequency $g$, and the classical laser pulse couples atomic states $\ket{e}$ and $\ket{s}$ with Rabi frequency $\Omega(t)$. 
In a rotating frame, the Hamiltonian within the rotating-wave approximation is (we set $\hbar=1$ throughout most of the manuscript) 

\begin{align}
H=&\> -\Delta_2  \ketbra{e}{e} + (\Delta_{1} -\Delta_2 )\ketbra{s}{s} \label{Hcoherent}
\\&+  (ga \ketbra{e}{g}+\Omega \ketbra{e}{s} + \text{h.c.}), \notag
\end{align}
where $a$ denotes the annihilation operator for the cavity field mode. Here $g=-\bra{e} \bm{d}\cdot \bm{\epsilon}_c \ket{g} \mathcal{E}_c/\hbar$ 
denotes the single-photon Rabi frequency, where $\mathcal{E}_c$ and $\bm{\epsilon}_c$ are the amplitude and the polarization of the single-photon electric field, respectively, and $\bm{d}$ is the atomic dipole moment operator. Similarly, $\Omega = -\bra{e}\bm{d}\cdot \bm {\epsilon}_p \ket{s}\mathcal{E}_p /2\hbar$ is the classical Rabi frequency, where $\mathcal{E}_p$ and $\bm{\epsilon}_p$ are the amplitude and the polarization of the classical laser pulse, respectively. Here, h.c.~denotes the Hermitian conjugate of the preceding two terms. The detunings $\Delta_1$ and $\Delta_2$ are defined by $\Delta_1=\omega_1-\omega_{es}$ and $\Delta_2 = \omega_c-\omega_{eg}$, respectively, where $\omega_1$ is the control laser frequency, $\omega_c$ is the cavity field frequency, $\omega_{es}$ is the frequency of the $e$--$s$ transition, and $\omega_{eg}$ is the frequency of the $e$--$g$ transition. This Hamiltonian is realized for both perfectly chiral and perfectly non-chiral polarizations 
of the cavity modes, as discussed in Sec.~\ref{sec:vqed}. In the perfectly chiral case, the transition dipole moment is oriented such that the cavity electric field enables a $\sigma^{\pm}$ transition between the atomic states $\ket{g}$ and $\ket{e}$. In the perfectly non-chiral
case, wherein the atom couples equally strongly to both clockwise and counterclockwise modes of the resonator ($a$ and $b$), the atom-cavity interaction is given by \cite{parkins}  
\begin{align}
H_{\text{int}}=(g a +g^* b) \ketbra{e}{g} + \text{h.c.}, \label{nonchiralH}
\end{align}
which can be reduced to Eq.~(\ref{Hcoherent}) by redefining $\frac{1}{\sqrt{2}}(e^{i\phi}a + e^{-i\phi} b) \rightarrow a$, where $g=\abs{g}e^{i\phi}$, and then redefining $\sqrt{2} g \rightarrow g$. After absorbing the phase of $g$ in the cavity mode operator ($ \propto  e^{i\phi}a+e^{-i\phi}b$), we can take $g$ to be real.
In this case, the transition dipole is oriented such that the cavity field enables a $\pi$ transition. We will therefore focus on Eq.~(\ref{Hcoherent}).
Additionally, we restrict our analysis to the radiatively limited case, where both optical coherences decay with rate decay $\gamma$, determined purely by the spontaneous emission rate of state $\ket{e}$. Our analysis can be extended to include additional dephasing of the $g$--$e$ and $g$--$s$ coherences (by adding additional decay rates to Eqs.\ (\ref{eom2}) and (\ref{eom1}), respectively).
We leave the problem of implementing single-photon retrieval and absorption in resonators with arbitrary polarization for future work.

Coupling the atom to the cavity allows for the use of the laser pulse $\Omega(t)$ to drive a Raman process that can generate or absorb single photons \cite{axel0, strp}. For single-photon generation, the system starts in state $\ket{s, 0}$ and is driven to state $\ket{g, 1}$ 
via a Raman process, where $\ket{x, n}$ denotes the joint atom-cavity state with the atom in state $x$ and the cavity in the $n$-photon number state. The cavity photon ($1$ in $\ket{g,1}$) is emitted into the waveguide.
Similarly, for single-photon detection, the input photon in the waveguide populates  state $\ket{g,1}$. The control pulse enables a Raman transition that drives state $\ket{g,1}$ to state $\ket{s,0}$ with $\ket{e,0}$ as the intermediate state. Measurement of the atomic state (e.g.,~using classical laser pulses \cite{Biedermann09}) enables detection of the photon. 

Using the method from Ref.~\cite{gauss}, we can compute, exactly, the control pulse $\Omega(t)=\Omega_0(t)e^{i\phi_0(t)}$ (for arbitrary detuning $\Delta_{2}$) required to generate a single photon with (normalized) mode shape $h(t)$, time duration $T$, and frequency
$\omega_c+\Delta_p$, such that 
$a_\text{out}=\sqrt{2\kx}c_g(t)=\sqrt{\eta_\text{r}}h(t)e^{-i\Delta_p t}$, where, as shown in Fig.~\ref{mt1}, $a_{\text{out}}$ is the mode corresponding to the outgoing photon and is normalized as $\int dt\> \abs{a_\text{out}}^2=\eta_{\text{r}}$. 
Here $c_g$ is the amplitude of the state $a^\dagger \ket{g, 0} = \ket{g,1}$, and $\eta_{\text{r}}$ is the efficiency of the single-photon generation (i.e.\ retrieval efficiency). Assuming that we choose $\Delta_1=\Delta_2+\Delta_p$, the phase $\phi_0$ is a slowly varying function. The exact expressions for $\Omega_0(t)$ and $\phi_0(t)$ are given in Appendix \ref{pulseshape}. To find analytic expressions 
for the required control Rabi frequency and the protocol's maximum efficiency $\eta_{\text{max}}$, we consider the adiabatic limit ($\kx T \gg 1$) to get
\begin{align}
\Omega_0(t)=&\frac{\xi}{g}\frac{A(t)}{[1-(\eta_{\text{r}}/\eta_{\max})\int_0^t h^2 dt^\prime]^{1/2}} , \label{omegat}\\
\xi^2=&(g^2-\Delta_p^2 -\Delta_2 \Delta_p + \kappa \gamma)^2  \label{scal}\\&\>+ [-\kappa(\Delta_p + \Delta_2)-\gamma \Delta_p]^2, \notag\\
A(t)=&\sqrt{\frac{\eta_{\text{r}}}{2\kx}}h(t) ,\\
\eta_{\text{max}}=&\frac{\kx}{\kappa}\Bigbl{1+C^{-1}+\frac{\gamma}{\kappa} \Bigb{\frac{\Delta_p}{g}}^2 }^{-1}, \label{effc}
\end{align}
where $2\gamma$ is the decay rate of the state $\ket{e}$, $\kappa=\kappa_{\text{i}}+\kappa_{\text{ex}}$, and $C=g^2/\kappa \gamma$ is the cooperativity parameter.
The corresponding expression for the phase of the control pulse $\phi_0(t)$ can be found in Eqs.~(\ref{phiexp1}--\ref{phiexp3}) in Appendix \ref{pulseshape}. In Eqs.~(\ref{omegat}--\ref{effc}), $\eta_{\text{r}}$ is a free parameter that can be chosen from zero to a value arbitrarily close to $\eta_{\text{max}}$, but the closer it is to $\eta_{\text{max}}$, the larger the value of the control pulse becomes near $t=T$. Moreover, numerics show that the exact profile for $\Omega_0(t)$ (computed without the adiabatic condition) for a Gaussian-like mode is non-divergent near $t=T$ even as $\eta_{\text{r}}$ becomes arbitrarily close to $\eta_{\text{max}}$ as long as we keep increasing $\kx T$ as we take $\eta_{\text{r}}$ closer to $\eta_{\text{max}}$
(the divergence here corresponds physically to the depletion of the amplitude in state $\ket{g,1}$). In practice, we can either choose $\eta_{\text{r}}=\eta_{\text{max}}$ and truncate the control pulse near $t=T$ or choose $\eta_{\text{r}} < \eta_{\text{max}}$ and avoid the truncation.
We remark that outside the limit $\kx T \gg 1$, we
can compute $\eta_{\text{max}}$ and the corresponding control pulse using numerical methods like gradient descent \cite{ascent}.

We also note that here we calculate the control pulse that (without truncation) retrieves a single photon exactly into the desired mode $h(t)$ with efficiency $\eta_{\text{r}}$, while the remaining fraction $1-\eta_{\text{r}}$ gets scattered via $\gamma$, lost via $\kappa_{\text{i}}$, or remains in $\ket{s}$. For many application purposes, a more relevant calculation would instead solve for the control pulse that maximizes the overlap of the output mode with the target mode $h(t)$, without asking for the output mode to be exactly $h(t)$. Such a calculation is difficult to do analytically since it requires an expression for $a_{\text{out}}$ in terms of the control pulse $\Omega(t)$ in the strong-coupling regime. At the same time, we expect that such a modified calculation will typically not give a qualitative improvement in the overlap efficiency compared to the optimization we use.

We now wish to compute the reduction of efficiency below $\eta_{\text{max}}$
due to insufficient control pulse power instead of fixing the value of $\eta_{\text{r}}$ and assuming unlimited control pulse power as in Eqs.~(\ref{omegat}--\ref{effc}). We can compute the retrieval efficiency in the limit $\kx T \gg 1$ as follows:
\begin{align}
\eta_{\text{r}}=\eta_{\text{max}}(1-e^{-2\kappa_{\text{ex}}g^2 f(0, T)/(\eta_{\text{max}}\xi^2)}) \label{effr},
\end{align}
where
\begin{align}
f(t, t^{\prime })&=\int_t^{t^{\prime }} \Omega_0^2(t^{\prime \prime})dt^{\prime \prime} .
\end{align}

Note here that the quantity in the exponent of Eq.~(\ref{effr}) depends on the frequency of the single photon retrieved [through $\xi$ in Eq.~(\ref{scal})] and therefore determines how large the deviation deviation of $\eta_\text{r}$ from $\eta_\text{max}$ is for a given value of $f(0, T)$. For the rest of the manuscript, we consider the case $\Delta_2=0$, i.e.\ when the cavity is resonant with the $e$--$g$ transition. Assuming $C \gg 1$, the quantity in the exponent simplifies to
\begin{align}
\frac{g^2 f(0, T)}{\xi^2}&= \frac{f(0, T)} {g^2 R(x)}\label{he1},\\
R(x)&= (1-x^2)^2+x^2(\kappa^2 + \gamma^2)/g^2  + 2C^{-1},  \label{he2}
\end{align}
where $x=\Delta_p/g$. In the following subsections, we will make use of this expression to analyze the dependence of efficiency on the available control power.

In single-photon absorption, we start with the state $\ket{g, 0}$ with an incoming single photon in the waveguide and end in the state $\ket{s, 0}$. This process corresponds qualitatively to the time-reversed version of the retrieval (single-photon generation) process. However, since the retrieval process ends with a small amplitude in the state $\ket{s, 0}$ (with an outgoing photon in the waveguide) while the absorption process starts with all of the amplitude in the state $\ket{g, 0}$ (with an incoming photon in the waveguide), the two processes are not exactly time-reversal symmetric. Numerical results show that the time-reversed version of the retrieval protocol (i.e., $\Omega(t)\rightarrow -\Omega(T-t)^*$) 
with retrieval efficiency $\eta_{\text{r}}$ gives approximately the same absorption efficiency $\eta_{\text{abs}} = |c_s(T)|^2$ as the retrieval efficiency, where $c_s(t)$ is the amplitude of the state $\ket{s, 0}$ at time $t$, and $T$ is the single-photon duration.

We will now study single-photon generation (i.e.\ retrieval) in two cases: in case 1, we will maximize the efficiency for unlimited control power, while, in case 2, we will maximize the efficiency for limited control power.
\subsubsection*{Case 1: Maximizing the efficiency for an unlimited control power}

In case 1, we consider maximizing the retrieval efficiency assuming unlimited control power. We suppose that the adiabatic condition $\kx T \gg 1$ is satisfied, so that Eqs.~(\ref{omegat}--\ref{effc}) hold.
To maximize $\eta_\text{max}$ in Eq.~(\ref{effc}), we set $\Delta_p=0$ to get the maximum efficiency
\begin{align}
\eta_{\text{max}}^{(1)}=\frac{\kx}{\kappa}[1+C^{-1}]^{-1}. \label{case1etam}
\end{align}
The retrieval efficiency can then be written as
\begin{align}
\eta_{\text{r}}=\eta_{\text{max}}^{(1)}\Bigb{1-e^{-2\kappa f(0, T)[1-C^{-1}]/g^2}}. \label{etaropt}
\end{align}
We now consider how the single-photon duration $T$ scales with the maximum control pulse size $\Omega_0$ and other physical parameters. We fix a value of $\eta_{\text{r}}$ close to $\eta_{\text{max}}^{(1)}$ in Eq.~(\ref{etaropt}) and find the following scaling:
\begin{align}
\Omega_0^2 \sim \frac{g^2}{\kappa T} \label{tc1},
\end{align}
which implies that, for a large fixed value of $\kappa T$, $\Omega_0 \sim g$. 
In the case $\kappa_{\text{i}}=0$ ($\kappa = \kx$), the leading error term for single-photon retrieval and absorption scales favorably as $C^{-1}$.
In the case where $\kappa_{\text{i}}$ is non-zero, the leading error term (ignoring higher orders in $C^{-1}$) is instead $\kappa_{\text{i}}/\kappa + (\kx/\kappa)C^{-1}$. We note here that, while choosing $\kx = \sqrt{g^2 \kappa_{\text{i}}/\gamma}-\kappa_{\text{i}}$ minimizes the error, this choice is only practical as long as $\kx$ is large enough to satisfy both $\kx T \gg 1$ and $T \ll \tau$.
We remark here that, based on Eq.~(\ref{tc1}), both $g$ and $\kx T$ determine how large the required control Rabi frequency $\Omega_0$ is, and therefore how feasible the scheme is. 

\begin{table*}[t]
\centering
\caption{Examples of various system parameters for case 1 (maximizing the efficiency for unlimited control power), where $g$ is the single-photon Rabi frequency, $\kx$ is the rate at which the microcavity couples to the waveguide, $\kappa_{\text{i}}$ is the microcavity's intrinsic loss, $2\gamma$ is the decay rate of the state $e$, $C=g^2/ \kappa \gamma$ 
is the cooperativity ($\kappa=\kx+\kappa_{\text{i}}$), and $T$ is the single-photon duration. The photons have frequency $\omega_c$. 
The angular frequencies and times have units of $(2\pi)$GHz and ns, respectively. $\mathcal{F}$ is the single-photon fidelity, and $\mathcal{F}_{\text{en}}$ is the atom-photon entangling  gate fidelity. Here $\eta_{\text{abs}}$ is the probability of measuring the state $s$ following the absorption of the single photon, and $\eta_d$ is the probability of detecting the correct atomic state using the controlled-phase-gate-based single-photon detection. Increasing the cavity length decreases $g$ and increases the transit time $\tau$. Parameters of cavities 1a, 1b, and 1c correspond to silicon nitride microdisk optical resonators~\cite{Barclay}; the parameters of cavities 2a and 2b correspond to higher refractive index silicon carbide microdisk optical resonators~\cite{Radulaski}; and the parameters of cavities 3 and 4 correspond to silicon nitride photonic crystal cavities in Ref.~\cite{alaien}.}
 \begin{tabular}{p{2.3cm} p{1.7 cm} p{1cm} p{1 cm} p{0.8 cm} p{1.1 cm} p{1cm} p{0.7 cm} p{1.2cm} p{1  cm} p{1 cm}p{1 cm}p{1 cm} p{0.6 cm} p{0.6 cm}}
\toprule
Setup & & $g$& $ \kx$ & $\kappa_{i}$ & $2\gamma$  & $C$ &  $\Omega_0$   & $1-\mathcal{F}$ & $1-\mathcal{F}_{\text{en}}$ & $1-\eta_{\text{abs}}$&$1-\eta_{d}$ &$T$ & $\tau$ & $\tau/T$ \\ \midrule 
Model cavities & Cavity 1a &  1.6 & 5 & 0.01& 0.0061  &  167 & 0.4 & 0.024 &  0.18 & 0.024 & 0.17& 2.22 & 13.7 & 6 \\
& Cavity 1b &  1.6 & 5 & 0.05 & 0.0061 & 166 & 0.4 & 0.032 & 0.19& 0.032 &  0.19& 2.22 & 13.7 & 6\\
& Cavity 1c &  1.6 & 3 & 0.01& 0.0061  &  280 & 0.5 & 0.02 &  0.05 & 0.02 & 0.04& 3.40 & 13.7 &  4 \\
& Cavity 2a &  4.7 & 7 & 0.01& 0.0061  &  1030 & 2.0 & 0.004 &  0.026 & 0.004 & 0.011 & 1.30 & 5.20 &  4 \\
& Cavity 2b &  2.0 & 6 & 0.01& 0.0061  &  220 & 0.6 & 0.025 &  0.18 & 0.025 & 0.17& 1.70 & 13.7 &  8 \\
 &Cavity 3 & 17 & 27 & 0.15 & 0.0061 & 3500 &7.0&  0.010 & 0.062 & 0.010& 0.031 & 0.24 & 1.3  &5 \\
 &Cavity 4 & 8 & 20 & 0.15 & 0.0061 & 1040  &3.0&  0.011 & 0.14 & 0.011& 0.12 & 0.43 & 1.3  &3 \\
 \midrule
\bottomrule
\end{tabular}
\label{tab1}
\end{table*}

In Table \ref{tab1}, we present results for exemplary parameter regimes wherein we can achieve both $C \gg 1$ and $\kx T \gg 1$ (and hence high efficiencies for single-photon sources and detectors) and also satisfy $T \ll \tau$.
Here the single-photon fidelity $\mathcal{F}$ is defined as $\mathcal{F}=\bra{\psi }\rho \ket{\psi }$ where $\rho$ is the output state and $\ket{\psi}$ is the target state. For model cavities in Table \ref{tab1}, parameters of cavities 1a, 1b, and 1c correspond to silicon nitride microdisk optical resonators~\cite{Barclay}, and parameters of cavities 2a and 2b correspond to higher refractive index silicon carbide microdisk optical resonators~\cite{Radulaski}. The parameters in cavities 3 and 4 correspond to silicon nitride photonic crystal cavities in Ref.~\cite{alaien} with mode volumes on the order of $\sim (\lambda/n_r)^3$, where $\lambda$ is the free-space wavelength, and $n_r$ is the material's refractive index. Here, values of $\kappa_{\text{i}}$ are consistent with theoretically predicted quality factors in such cavity types, and in some cases, such values have been approached experimentally. The values of $\kx$ correspond to cases wherein the cavities have strong overcoupling ($\kx \gg \kappa_{\text{i}}$) 
with a bus waveguide. The values of the classical Rabi frequency correspond to optical intensities ($\sim 10^3 q^2$ W cm$^{-2}$, where $q$ is the size of the Rabi frequency $\Omega_0$ in GHz, for the atomic level scheme proposed in Sec.~\ref{atomiclsch}) achievable on-chip because laser power in a waveguide is confined to sub-$\mu $m$^2$ area.

We first note that, since we have high values of $g$, we can choose $\kx > g $ so that the single-photon duration $T$ is small while keeping the cooperativity large (as in cavity 1a). In general, this helps us maximize the number of single-photon pulses we can generate during an atomic transit. We first examine the role of intrinsic loss $\kappa_\text{i}$ by comparing cavity 1b to cavity 1a and see that, even in this regime where $C$ is only marginally impacted, larger $\kappa_{i}$ causes a reduction in single-photon fidelity. As shown in model cavity 1c, we can also decrease $\kx$ in favor of a larger cooperativity. However, this leads to larger values of single-photon duration to satisfy $\kx T \gg 1$, and hence a lower number of photons produced per transit event. 
We can also decrease the value of $g$ in favor of a longer cavity length and a longer transit time, as shown in model cavities 2a and 2b. However, this requires reduced values of $\kx$ accordingly to maintain high cooperativity, which also means we will need larger values of single-photon duration $T$ to satisfy $\kx T \gg 1$. Finally, cavities 3 and 4 show performance with a very limited transit time (e.g., in the limit of no beam collimation), where large $g$ and $\kx$ can enable a few high-fidelity operations to be performed. In Appendix \ref{ultracold}, we study how the aforementioned performance metrics are enhanced by using various atomic cooling techniques.

\subsubsection*{Case 2: Maximizing the efficiency for a limited control pulse size}

\begin{figure}[t]
		\includegraphics[
		width=\columnwidth
		]{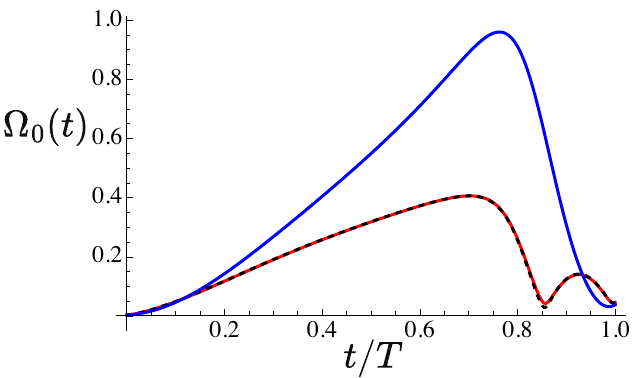}
	\caption{Size of control pulse $\Omega_0(t)$ [in $(2\pi)$GHz] used to retrieve a single photon with mode shape $\propto \sin^2(\pi t/T)$ with system parameters $(g,\kx, \kappa_{\text{i}}, 2\gamma )=2\pi(1.6, 0.9, 0.01, 0.0061)$GHz. The blue line is for case 1 ($\Delta_p=0$) with efficiency $\eta_{\text{r}}=0.979$. The red line is for case 2 with the value of $\Delta_p=-x_1 g$ obtained through numerical maximization of $\eta_{\text{r}}$ in Eq.~(\ref{effr}) w.r.t.~$\Delta_p$ with $f(0, T) = 1.34g$  achieving $\eta_r = 0.979$. The dashed black curve is for case 2 with the value of $\Delta_p=-x_0g$ with the same value of $f(0, T)$. As seen from Eq.~(\ref{etaropt}), $f(0, T)/(g^2/2\kappa)=1.53$ shows that photons cannot be retrieved (with high efficiency) for this value of $f(0, T)$ at the cavity resonance frequency. Moreover, as seen from Eq.~(\ref{effc2}), $f(0, T)/((\kappa+\gamma)^2/(2\kx))=8.84 \gg 1$, which explains why photons can be retrieved at frequency $\Delta_p = -x_0 g$ with high efficiency.} 
	\label{sizecomp1}
\end{figure}

We now consider case 2, i.e., when we have a limited control pulse size and $\kx T \gg 1$. We can then maximize $\eta_{\text{r}}$ in Eq.~(\ref{effr}) with respect to $x=\Delta_p /g$ for a given value of $f(0, T)$. 
However, instead of maximizing the full expression for $\eta_\text{r}$ (which depends on $\eta_\textrm{max}$ and the value of the exponential), we note that minimizing just the value of the exponential for the parameters considered gives approximately the same efficiency. This is because, for limited $f(0, T)$, the retrieval efficiency is more sensitive to a non-negligible value of the exponential than to a smaller value of $\eta_{\text{max}}$ (which is always close to one for $C \gg 1$, $\kappa_{\text{i}} \ll \kx$, $\gamma/\kappa \ll 1$, and $|x|<1$). Then, for $ \kappa^2 + \gamma^2 < 2g^2$, Eqs.~(\ref{effr}$-$\ref{he2}) show that we can choose the single-photon frequency $\Delta_p=x_0 g$ (where $\pm x_0$ are the minima of $R(x)$ in Eq.~(\ref{he2})) such that
\begin{align}
x_0&=\pm \sqrt{1-\frac{\kappa^2+\gamma^2}{2g^2}}, \label{defx0}\\
\Delta x & \sim \frac{2\kappa}{\sqrt{4g^2-\kappa^2}}, \label{widx}
\end{align}
where $\Delta x$ is the width around the minima of $R(x)$.
The correction to the efficiency due to finite control pulse power (in the limit $\kx T \gg 1$) can then be computed from Eq.~(\ref{effr}) as follows: 
\begin{align}
\eta_{\text{r}}=\eta_{\text{max}}^{(2)}(1-e^{-2\kappa_{\text{ex}}f(0, T)/(\eta_{\text{max}}^{(2)} (\kappa  + \gamma )^2)  }), \label{effc2}
\end{align}
where the maximum efficiency is now
\begin{align}
\eta_{\text{max}}^{(2)}=\frac{\kx}{\kappa}\Bigbl{1+C^{-1}+x_0^2\frac{\gamma}{\kappa}}^{-1}, \label{etamaxc2}
\end{align}
where we note that the maximum achievable efficiency is smaller than the one from case 1 in Eq.~(\ref{case1etam}). Moreover, we get the following scaling for $T$:
\begin{align}
\Omega_0^2 \sim \frac{(\kappa + \gamma)^2}{\kappa T}, \label{tc2}
\end{align}
implying that, for some large fixed value of $\kappa T$, $\Omega_0 \sim (\kappa+\gamma)$.
When we choose $x=x_0 \approx 1$ ($\Delta_p \approx g$), the state $\ket{+} \propto \ket{e, 0} + \ket{g, 1}$ is coupled resonantly to the state $\ket{s,0}$, which allows us to adiabatically eliminate the state $\ket{-} \propto \ket{e, 0}- \ket{g, 1}$. Physically, this corresponds to addressing one of the eigenstates of the atom-cavity Hamiltonian. This explains why we have a lower value of $\eta_{\text{max}}$ in case 2 compared to case 1, since the $\ket{e, 0}$ state gives an additional contribution to the decay. 
One advantage of case 2 is that we get a better scaling for $\Omega_0$ ($\Omega_0 \sim \kappa+\gamma$) compared to case 1 (where $\Omega_0 \sim g$) in that we need smaller control pulse sizes since $(\kappa + \gamma) \ll g$ can be easily achieved in the strong coupling regime in experiment. However, 
one disadvantage of case 2 is that the photons retrieved cannot be used for implementing the atom-photon controlled-phase gate, as they are off-resonant (by approximately $\pm g$) from the cavity resonance frequency. Instead, in the limits $\kappa_{\text{i}} \ll \kappa_{\text{ex}}$, $g \gg \gamma$, the atom-cavity interaction implements the following atom-photon gate:
\begin{align}
W=Z_p \text{CP}(\phi),
\end{align}
where $Z_p=\text{diag}(1, -1)$ in the basis $\{\ket{v}, \ket{h} \}$, $\text{CP}(\phi)=\text{diag}(1, 1, 1, e^{i\phi})$ in the basis $\{\ket{gv}, \ket{gh}, \ket{sv}, \ket{sh} \}$, and $\phi = 2\pi - 2\arctan(g/\kappa_{\text{ex}})$. The derivation of this gate follows from Eq.~(\ref{tcf}). Since the value of $g$ fluctuates with the atomic position, we leave it as an open question whether a changing but known value of $\phi$ (obtained by monitoring the atomic position) in the CP$(\phi)$ gate can be used to perform universal quantum computation.  
In Fig.~\ref{sizecomp1}, we compare the control pulses $\Omega_0(t)$ for the two cases, $x=0$ (case 1) and $x=x_0$ (case 2).
In this example, for case 2, we also compare two different control pulses. The first control pulse is obtained from setting $x=x_0$, producing a single photon with the corresponding value of $f(0, T)=f_0$. The second control pulse is obtained from the value of $x$ that maximizes $\eta_{\text{r}}$ in Eq.~(\ref{effr}) for $f(0, T)=f_0$. The values of $x$ are close ($x_0=0.916$, while numerical maximization gives $x=0.894$).

We now consider the dominant errors for case 2. First, consider the situation where $\kappa_{\text{i}}=0$. In this case, Eq.~(\ref{effc}) shows that the dominant inefficiency is $x_0^2 (\gamma/\kappa) + C^{-1}$. In the situation where $\kappa_{\text{i}} \ll \kx$, Eq.~(\ref{effc}) shows that the dominant inefficiency is $(\kappa_{\text{i}} +x_0^2 \gamma)/\kx + (\kx/\kappa)C^{-1}$  

\begin{figure}[t]
	\begin{center}
	\includegraphics[
	width=\columnwidth
		]{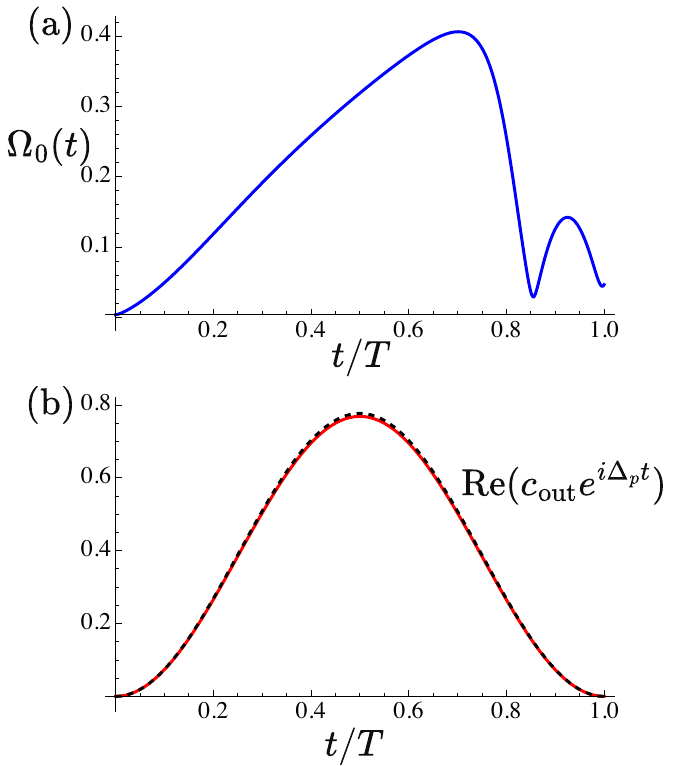}
	\end{center}
	\caption{(a) Size of the control pulse to retrieve a single photon in case 2 ($\Delta_p=-x_0 g$) with mode shape $\sin^2(\pi t/T)$ and with system parameters $(g,\kx, \kappa_{\text{i}}, 2\gamma )=2\pi(1.6, 0.9, 0.01, 0.0061 )$GHz with $T=4.42$ ns. (b) The red line corresponds to the real part of $a_{\text{out}}e^{i\Delta_p t}=\sqrt{2\kx}c_g e^{i\Delta_p t}$, and the dashed black line is the desired mode function $\propto \sin^2(\pi t/T)$. The inefficiency is $1-\eta_{\text{r}}=0.021$.}
	\label{c2ret}
\end{figure}

We can choose a value of $\kx$ based on experimental constraints, e.g.,~minimizing the control pulse size, or minimizing photon pulse duration $T$ with $\kx T \gg 1$, while keeping $\kappa$ small enough to be consistent with Eq.~(\ref{defx0}). In the examples considered in Table \ref{tab2}, we show parameters with near-ideal efficiencies and $T \ll \tau$. In the table, we also compare the maximum sizes of the control pulse needed to retrieve/absorb photons with given efficiencies using control pulses from case 1 and case 2. In Fig.~\ref{c2ret}, 
we show the retrieval of a single photon with system parameters from model cavity 5 in Table \ref{tab2}.

\subsection{Atomic level scheme} \label{atomiclsch}

\begin{table*}[htb]
\centering
\caption{Examples of various system parameters for case 2 (maximizing the efficiency for a limited control pulse size), where $g$ is the single-photon Rabi frequency, $\kx$ is the rate at which the microcavity couples to the waveguide, $\kappa_{\text{i}}$ is the microcavity's intrinsic loss, $2\gamma$ is the decay rate of state $\ket{e}$, $C=g^2/ \kappa \gamma$ is the cooperativity ($\kappa=\kx+\kappa_{\text{i}}$), $T$ is the single-photon duration, and $\Delta_p=-gx_0$. The angular frequencies and times have units of $(2\pi)$GHz and ns,  respectively. Here $\mathcal{F}$ is the single-photon fidelity, and $\eta_{\text{abs}}$ is the probability of measuring the state $s$ following the absorption of the single photon. Increasing the cavity length decreases $g$. The maximum size of the control pulse for case 2, $\Omega_0$, is compared with the maximum size of the control pulse from case 1, $\tilde{\Omega}_0$, used to retrieve/absorb single photons with the same parameters and frequency resonant with the cavity (i.e.\ $\Delta_p = 0$). Parameters of cavity 5 correspond to silicon nitride microdisk optical resonators~\cite{Barclay}, and the parameters of  cavities 6 and 7 correspond to silicon nitride photonic crystal cavities in Ref.~\cite{alaien}.}
 \begin{tabular}{p{3cm} p{2 cm} p{0.6cm} p{0.6 cm} p{0.8 cm} p{1.1 cm} p{1cm} p{1 cm} p{1 cm} p{1 cm} p{1.1  cm} p{1 cm} p{0.6 cm} p{0.6 cm}}
\toprule
Setup &  & $g$& $ \kx$ & $\kappa_{i}$ & $2\gamma$  & $C$ &  $ \Omega_0$ & $ \tilde{\Omega}_0$   & $1-\mathcal{F}$ & $1-\eta_{\text{abs}}$&$T$ & $\tau$ & $\tau /T$\\ \midrule 
Model cavities & Cavity 5 &  1.6 & 0.9 & 0.01& 0.0061  &  920 & 0.4 & 1.0 &  0.021  & 0.021 & 4.42 & 13.7 &3\\
& Cavity 6 &  17& 12 & 0.15 &   0.0061 & 7800 & 5.4
 & 12.0 &0.014   & 0.014 & 0.42 &1.3 & 3\\  
 & Cavity 7 &  8& 6 & 0.15 &   0.0061 & 3400 & 3.0
 & 6.0 &0.027   & 0.027 & 0.64 &1.3 & 2\\
\midrule
\bottomrule
\end{tabular}
\label{tab2}
\end{table*}

\begin{figure}[b]
	\begin{center}
		\includegraphics[
		width=\columnwidth
		]{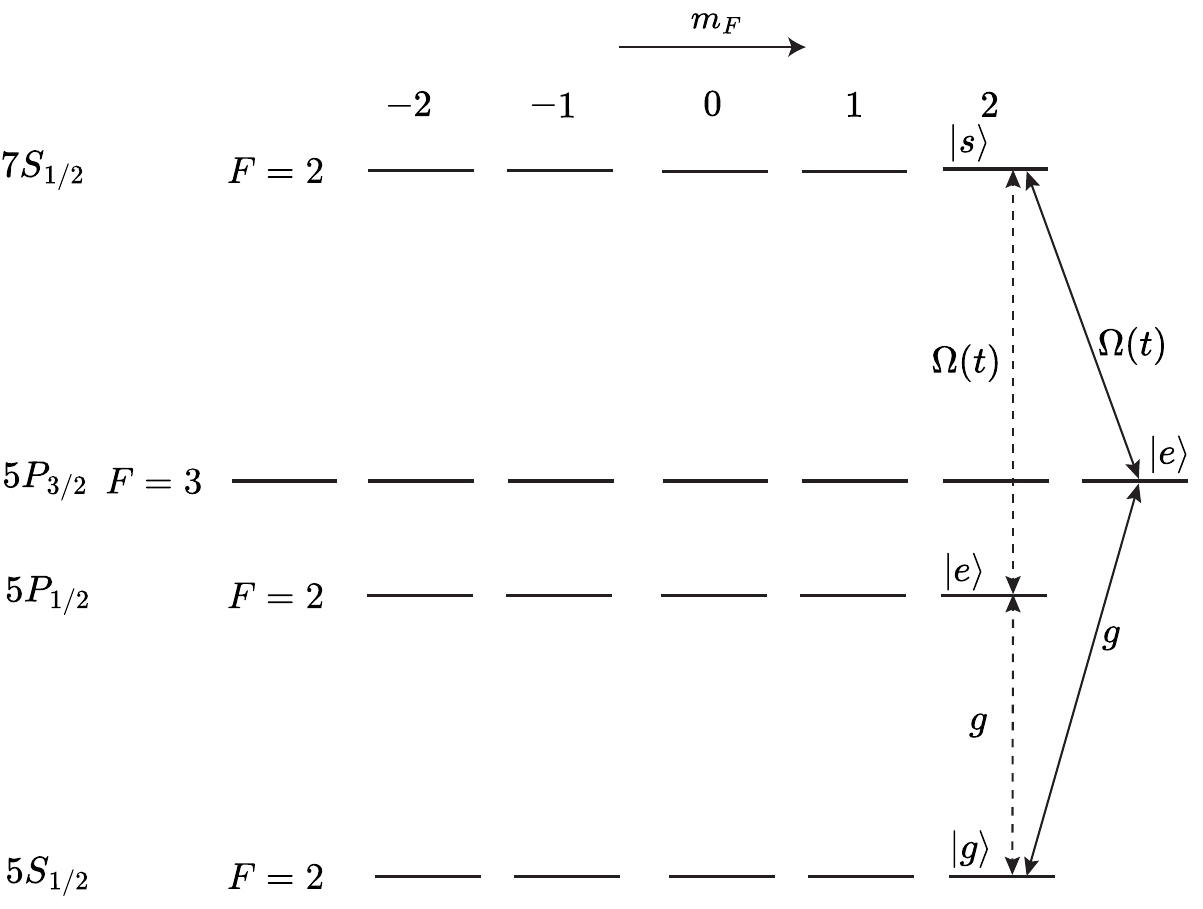}
	\end{center}
	\caption{Proposed atomic level scheme for the chiral case (solid black) and non-chiral case (dashed black).
    For the chiral case, mode $a$ couples to the $\sigma_+$ $g$--$e$ transition. The $s$--$e$ transition has wavelength 741 nm, and the $e$--$g$ transition has wavelength $780$ nm. For the non-chiral case, modes $a$ and $b$ couple to the $\pi$ $g$--$e$ transition. The $s$--$e$ transition has wavelength 728 nm, and the $e$--$g$ transition has wavelength $795$ nm. The detunings are defined as $\Delta_1=\omega_{se}-\omega_1$ and $\Delta_2=\omega_c-\omega_{eg}$, where $\omega_{ij}$ is the energy corresponding to the atomic transition $i\rightarrow j$, $\omega_1$ is the frequency of the control pulse $\Omega(t)$, and $\omega_c$ is the cavity resonance frequency.}
	\label{atomicLevelDF}
\end{figure}

We study the two cases i.e.,~the case where the optical modes are perfectly chiral, assuming the mode $a$ has perfect circular polarization, and the case where the optical modes are perfectly non-chiral, assuming the modes $a$ and $b$ have perfectly linear polarization. In the chiral case, the single photon is retrieved and absorbed into  mode $a_\text{out}$. 
However, in the non-chiral case, the single photon is retrieved and absorbed into the mode $ \propto (e^{i\phi}a + e^{-i\phi}b)$, where $\phi$ is the phase defined in  Eq.~(\ref{HamTE})]. In the case of retrieval, we can use the position measurement of the atom to determine the phase $\phi$, and then 
transform the retrieved mode into $a_{\text{out}}$ (or $b_{\text{out}}$) using linear optics. Similarly, for absorption, we must first use linear optics to transform the mode we want to be absorbed, say $a_{\text{in}}$, into mode $\propto (e^{i\phi}a_{\text{in}}+e^{-i\phi}b_{\text{in}})$, and then send it to the cavity (and apply the control pulse as described in Sec.~\ref{singlephoton}). 
We propose using the following atomic level schemes for the chiral and non-chiral cases 
(as shown in Fig.~\ref{atomicLevelDF}). 
For the chiral case, we choose $\prescript{87}{}{\text{Rb}}$ with $\ket{g}=\ket{5S_{1/2}, F=2, m_F=2}$, $\ket{s}=\ket{7 S_{1/2}, F=2, m_F=2}$, and $\ket{e}=\ket{5P_{3/2}, F=3 , m_{F}=3}$. The $s$--$e$ transition has wavelength $741$ nm, and the $e$--$g$ transition has wavelength 780 nm. For the non-chiral case, we choose $\prescript{87}{}{\text{Rb}}$  with $\ket{g}=\ket{5S_{1/2}, F=2, m_F=2}$, $\ket{s}=\ket{7 S_{1/2}, F=2, m_F=2}$, and $\ket{e}=\ket{5P_{1/2}, F=2, m_{F}=2}$. The $s$--$e$ transition has wavelength $728$ nm, and the $e$--$g$ transition has wavelength $795$ nm. 

\begin{figure}[t] 
\includegraphics[width=\columnwidth]{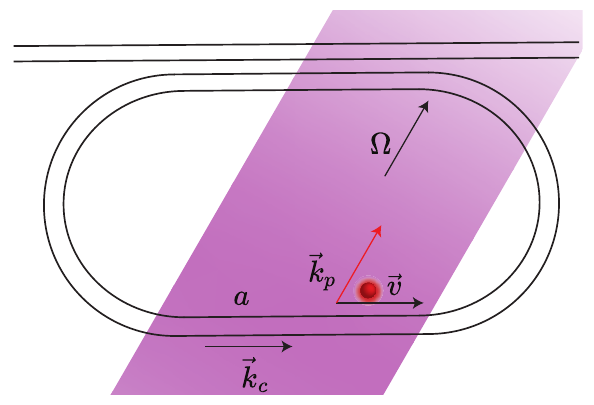}
\caption{Scheme for dealing with the Doppler shift. Shown here is a chiral cavity with optical mode $a$ with wave-vector $\vec{k}_c$. The control pulse $\Omega$ has the wave-vector $\vec{k}_p$ at an angle to the atom's velocity, which is parallel to the microcavity's length. The angle is chosen to cancel the unwanted contribution to two-photon detuning due to the Doppler effect. For a lambda system, the condition $\vec{k}_c.\vec{v} -\vec{k}_p.\vec{v}=0$ cancels the Doppler contribution to the two-photon detuning. For the ladder system (as in Fig.~\ref{atomicLevelDF}), this condition is modified to $\vec{k}_c.\vec{v} +\vec{k}_p.\vec{v}=0$.}
\label{cavsk}
\end{figure}

We now focus on the chiral case where we can ameliorate the effect of large atomic velocities on photon retrieval and absorption. Since the $e$--$g$ transition has a longer wavelength than the one corresponding to the $s$--$e$ transition,  applying the control pulse at an angle to the cavity, as shown in Fig.~{\ref{cavsk}}, can cancel the unwanted contribution to the two-photon detuning  
caused by the Doppler effect.
The chosen state $\ket{s}$ has a lifetime of $88$ ns \cite{Gomez_2004}, which is much larger than the values of single-photon duration $T$ we consider. While this atomic level scheme is not a lambda-type configuration (it is a ladder configuration), our equations still describe the system (as long as we redefine the detuning $\Delta_1$) since the state $\ket{e}$ still has the dominant decay rate. After redefining $\Delta_1$ to $\omega_{se}-\omega_1$ for the ladder configuration, our control pulse scheme can be used in the same way. The laser and cavity fields are chosen to satisfy $\vec{k}_p \cdot \vec{v}+ \vec{k}_c \cdot \vec{v} \approx 0$ (where $\vec{k}_p$ and $\vec{k}_c$ are the wave vectors corresponding to the control pulse and the cavity mode, respectively, and $\vec{v}$ is the velocity of the atom), minimizing the unwanted Doppler contribution to the two-photon detuning $\Delta_1 - \Delta_2$, as described in Fig.~\ref{atomicLevelDF}. The three-level-atom approximation holds because the unwanted off-resonant transitions have large detunings compared to the corresponding Rabi frequencies. For the chiral case, the cavity field is detuned from the $\ket{5D}$--$\ket{e}$ and $\ket{g}$--$\ket{5P_{1/2}}$  transitions by $ \approx (2\pi)2$ THz and $\approx (2\pi) 7$ THz, respectively, with single-photon Rabi frequencies comparable to $g \approx (2\pi) 10$ GHz. Similarly, the control pulse is detuned from the $\ket{s}$--$\ket{5P_{1/2}}$ transition by $\approx(2\pi)7$ THz with Rabi frequency comparable to $\Omega \approx (2\pi)10$ GHz. {For producing a string of single-photon pulses, we need to reinitialize the state to $\ket{s, 0}$ after each single-photon retrieval (which ends in the state $\ket{g, 0}$). The state can be reinitialized using a pair of $\pi$-pulses on the transitions $\ket{g}$--$\ket{5P_{1/2}}$ and $\ket{5P_{1/2}}$--$\ket{s}$.} Finally, we note here that, even when we can't use the same scheme for a non-chiral system, wherein it is impossible to cancel the Doppler shift to the two optical modes simultaneously for the geometry shown in Fig.~\ref{cavsk}, it is possible that a high atomic velocity may not be an issue. For instance, if the atom passes transversely to the cavity wave-vector, the detunings $\Delta_1$ and $\Delta_2$ will have no Doppler shift contributions.

We now compare the proposed sources and detectors with other platforms. The single-photon detection efficiencies obtained by the cavities discussed in Tables \ref{tab1} and \ref{tab2} compare well to efficiencies achievable by superconducting nanowire single-photon detectors which can have detection efficiencies of approximately 0.97 at 1500 nm (at $10^5$ counts/s) \cite{snspd}. We then compare the possible single-photon fidelities from our heralded  single-photon generation to other available platforms.  State of the art spontaneous parametric-down-conversion, quantum-dot, and Rydberg-ensemble  platforms can have heralded single-photon fidelities $\mathcal{F}$ of 0.961  \cite{spdcsrc1}, 0.972 \cite{qdsrcA}, and 0.982 \cite{atomsrcryd}, respectively, where $\mathcal{F}=\bra{\psi }\rho \ket{\psi }$, $\ket{\psi}$ is the target state, and $\rho$ is the output state. These are also comparable to single-photon fidelities achievable with the parameters we consider in Tables \ref{tab1} and \ref{tab2}.
For single-photon sources, the constraint on the number of photons produced per atomic transit (given approximately by $\tau/T$) arises from several factors, including the atomic transit time $\tau$, finding feasible values of $\kappa$, $T$ that satisfy $\kappa T \gg 1$, $C \gg 1$ and $T \ll \tau$, the feasibility of achieving sufficiently large Rabi frequencies that satisfy Eqs.~(\ref{tc1}) and (\ref{tc2}), and performing sufficiently fast atomic state rotations that reset the atomic state to the correct initial state.
Similarly, for single-photon detection, the constraint on the number of photons detected per transit event (given approximately by $\tau/T$, where $T$ is the incoming single photon's duration) comes from the atomic transit time $\tau$, single-photon duration $T$, and performing sufficiently fast atomic state rotations to reset the atomic state to the correct initial state.
We emphasize here that the probability of having a successful heralded single-photon-generation event differs for the above-mentioned parametric down-conversion, quantum-dot, and Rydberg-ensemble platforms. The Rydberg-ensemble and quantum-dot sources are deterministic in the sense that the atomic gas and the quantum dot are always present. However, single-photon generation from spontaneous parametric down-conversion has a small heralded success probability. Such heralded sources can be made near-deterministic using multiplexing \cite{spdcmux}. Similarly, we propose using active and passive multiplexing, as discussed in the following section, to make our single-photon sources and single-photon detectors (which both rely on heralding to check if the cavity is active, i.e.\ if it has an atom in it) near-deterministic.

\subsection{Multiplexing} \label{sec3A}
We now outline how previously discussed protocols for single-photon generation and detection can be multiplexed both actively (i.e.,~monitoring the cavities and subsequently applying feedback using control switch networks to route single photons) and passively (i.e.,without monitoring or feedback).
\subsubsection{Active multiplexing} \label{afirst}
\begin{figure}[t] 
\hspace*{1cm}
\includegraphics[width=0.85\columnwidth]{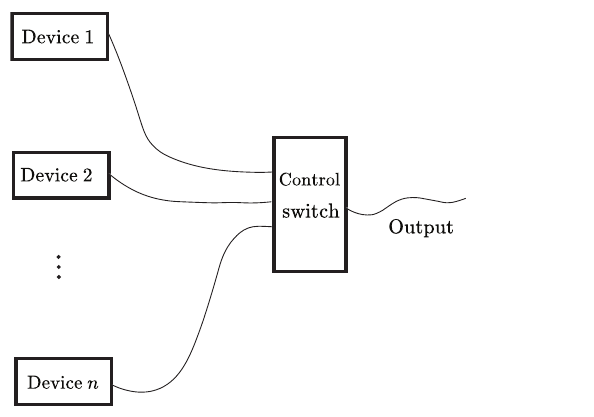}
\caption{Scheme for making a multiplexed single-photon source. The control switch in this case monitors which device is active and routes the single photon to the output. }
\label{source}
\end{figure}

\begin{figure}[t] 
\includegraphics[width=\columnwidth]{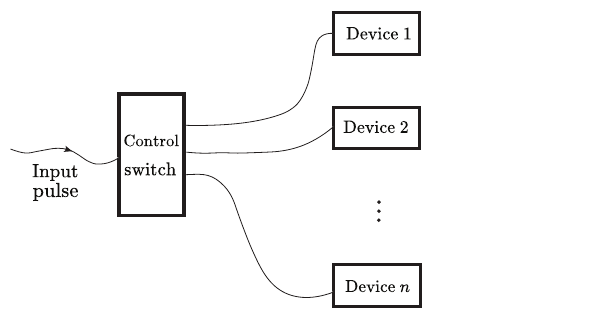}
\caption{Scheme for making a multiplexed single-photon detector.}
\label{detect}
\end{figure}

\begin{figure}[t] 
\includegraphics[width=1.0\columnwidth]{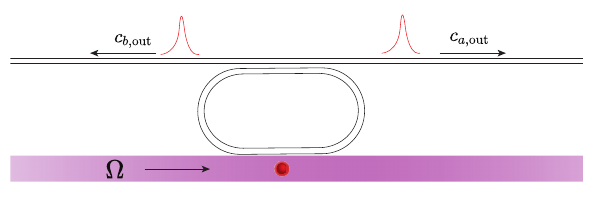}
\caption{Single-photon source in the non-chiral case. This requires monitoring the atom's position and the phase of the cavity field defining the optical mode.}
\label{MulChNC}
\end{figure}

For implementing an actively multiplexed single-photon source, several of our single-photon sources (which may be active or not active at any given time) are connected to a control switch as shown in Fig.~\ref{source}. Detection of an atom in the cavity before the protocol starts ensures that a heralded single photon is produced. It is also possible to herald the output photon by measuring the atomic state $\ket{g}$ at the end of the protocol. The switching time should be at least an order of magnitude smaller than the transit time i.e.,~at most in the ns range. Similarly, for implementing a multiplexed single-photon detector (as shown in Fig.~\ref{detect}), a control switch routes the input pulse to an active cavity. This scheme works for both the chiral and non-chiral systems. In the non-chiral case (as shown in Fig.~\ref{MulChNC}), active monitoring of the system is necessary to measure the phase $\phi$ that defines the optical mode $ \propto (e^{i\phi}a + e^{-i\phi}b)$.

\subsubsection{Passive multiplexing} \label{fpass}
\begin{figure}[t] 
\includegraphics[width=1.0\columnwidth]{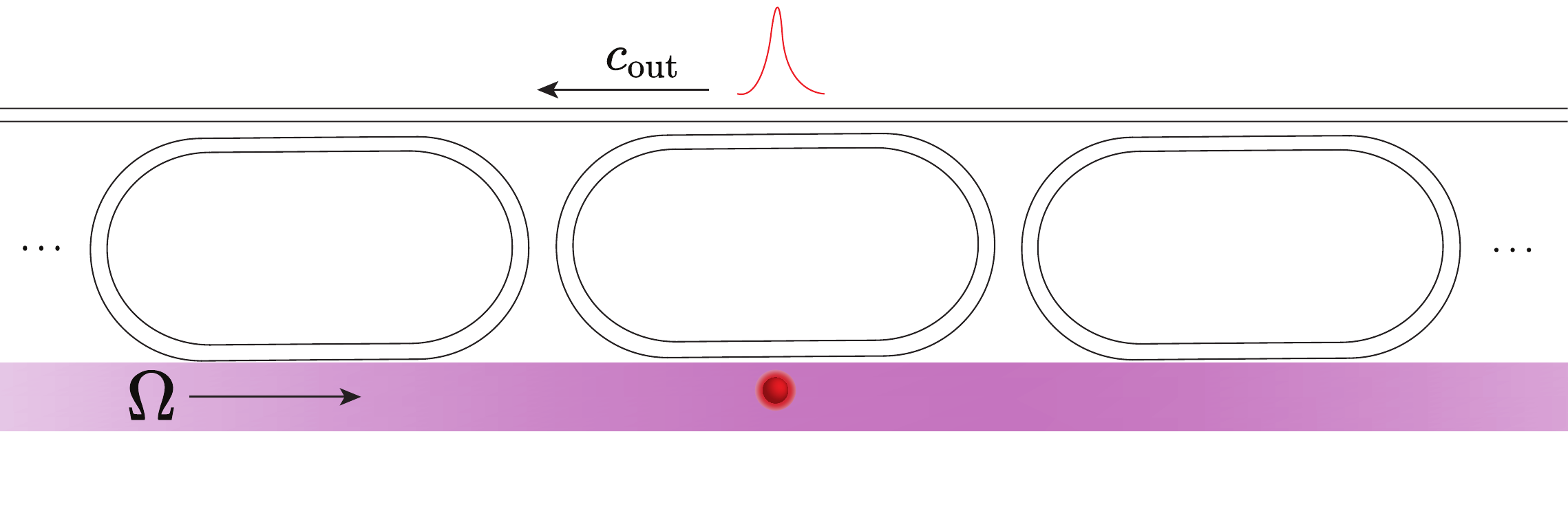}
\caption{Passively multiplexed single-photon source in the chiral case. The number of cavities is chosen to maximize the probability of having exactly one active cavity. The position of the atom is not monitored.}
\label{MulCh}
\end{figure}

As shown in Fig.~\ref{MulCh}, we use an array of $N$ cavities to maximize the probability that exactly one of them is active for implementing a passively multiplexed source. For the retrieved single-photon wavefunction to be independent of which cavity is active, we can apply a different control pulse with a suitable time delay to each cavity. 
We now consider the case where $p$ is the probability that a single cavity is active and assume that this probability is independent of whether other cavities are active or not. Then, maximizing the probability $\tilde{p}$ that exactly one cavity is occupied gives us the optimal number of cavities $\tilde{N}=-1/\ln(1-p) \approx 1/p -1/2 - p/12 + O(p^2)$ with the corresponding value  $\tilde{p}=e^{-1}(1+p/2)+O(p^2)$. However, for $\kappa_{\text{i}} \neq 0$, the single photon experiences photon loss as it passes through the $N$ cavities, leading to an inefficiency contribution that scales, roughly, as $N \kappa_{\text{i}}/\kx$ [see Eq.~(\ref{tcoeff})]. This scheme works for chiral cavities where the phase measurement of the cavity field is not required. 

\subsection{Experimental considerations} \label{expgf}
We now consider the experimental conditions necessary to achieve near-ideal single-photon generation and absorption. We consider two non-idealities specific to warm atoms: temporal variation in the coupling $g$ and the Doppler shift of the frequencies of the control pulse and the cavity field. We ignore transit-time broadening because we consider schemes with $T \ll \tau$. 

\begin{figure}[tp]
    \includegraphics[
		width=\columnwidth
    		]{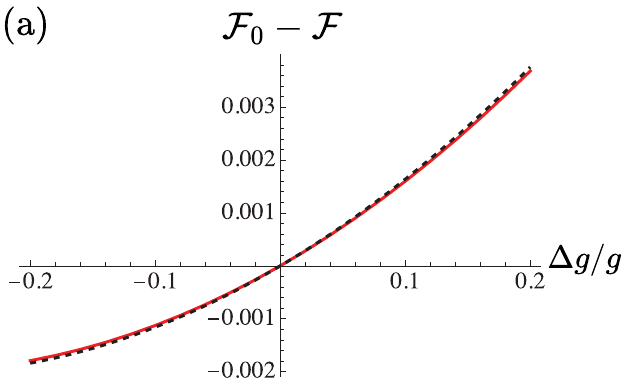} \includegraphics[
		width=\columnwidth
    		]{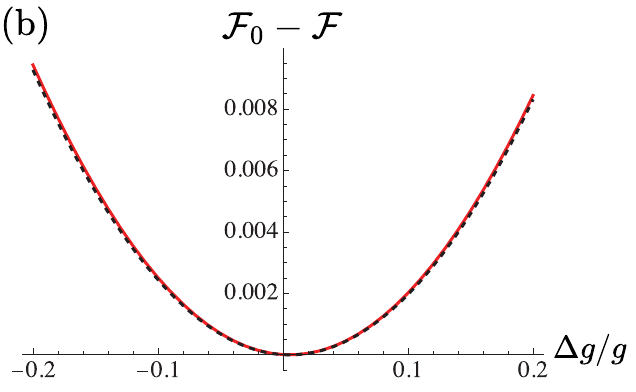}
	\caption{We plot here the correction to the fidelity of the retrieved photon due to temporal variation in $g$, as described in Sec.~\ref{expgf}.
    (a) The red line corresponds to parameters in model cavity 1a (for case 1 in Table \ref{tab1}). The correction is described, approximately, by the fit 0.014($\Delta g/g$)+0.024$(\Delta g/g)^2$ (shown by the black-dashed line). (b) The red line corresponds to the correction to the fidelity of the retrieved photon for parameters in model cavity 5 (for case 2 in Table \ref{tab2}). The corrections are described, approximately, by the fit $-0.0024 (\Delta g/g)+0.22(\Delta g/g)^2$ (shown by the black-dashed line).}
	\label{retfidc1}
\end{figure}

We first consider the effect of the variation in time of the coupling strength $g$. We suppose that, during the time from $t = 0$ to $t = T$, the coupling $g(t)$ varies so little that we can expand $g(t)$ to first order in $t$ as follows: $g(t)=g + \Delta g (t-T/2)/T$. Here $\Delta g \sim (g/L)(v_\perp T)$, where $v_\perp$, the component of the atom's velocity perpendicular to the cavity, is determined by the atomic beam divergence, and the cavity length $L$ is the typical length scale over which $g$ varies. To study the effect of the variation in $g(t)$ on retrieval, we attempt to retrieve into mode $\propto \sin^2(\pi t/T)$ using a control pulse from Eqs.~(\ref{controlP}, \ref{controlphi}) designed assuming that $g(t) = g$. We then compute the infidelity defined as $1-\bra{\psi}\rho \ket{\psi}$, where $\rho$ is the output state and $\ket{\psi}$ is the target state. For absorption, we attempt to absorb the mode $\propto \sin^2(\pi t/T)$ using a time-reversed version of the control pulse from Eqs.\ (\ref{controlP}, \ref{controlphi}) designed assuming that $g(t) = g$. We then compute the inefficiency defined as $1-\abs{c_s(T)}^2$, where $|c_s(T)|^2$ is the probability of the atom being in state $s$ at $t=T$. For case 1 parameters in Table \ref{tab1}, numerics show that, for retrieval, the contribution to infidelity due to variation in $ g$, $\mathcal{F}_0-\mathcal{F}$, where $\mathcal{F}_0 (\mathcal{F})$ is the single-photon fidelity for the case $\Delta g=0$ ($\Delta g \neq 0$), is well-approximated by $a_1 (\Delta g/g) + a_2 (\Delta g/g)^2$ with $\abs{a_1} < \abs{a_2}$. For the example shown in Fig.~\ref{retfidc1}(a), we have $(a_1, a_2) =(0.014, 0.024)$.  
For case 2, since the resonance about the optimal point $x_0$ has width $\Delta x$ [defined in Eq.~(\ref{widx})], the error depends on the relative size of $(\Delta g/g)$ and $\Delta x$. In particular, one can use a higher $\kx$ to broaden the resonance and accommodate a larger error $\Delta g$. Numerics show that, as long as the condition $ |\Delta g/g| < \Delta x$ holds, then for retrieval, the contribution to the infidelity due to $\Delta g$ is well-approximated by $d_1 (\Delta g/g)+d_2(\Delta g/g)^2$ with $\abs{d_1} \ll \abs{d_2}$. For the example in Fig.~\ref{retfidc1}(b), $(d_1, d_2) = (-0.0024, 0.22)$. The smaller $\Delta g/g$ is relative to $\Delta x$, the smaller the value of $d_2$. 
In cases 1 and 2, inefficiencies due to temporal variation of $g$ for single-photon absorption obey similar error scalings as the corresponding inefficiencies for single-photon generation. As an example of experimentally relevant values of $\Delta g/g$, using the parameters of model cavity 2a with an atomic beam divergence of $0.013$ rad \cite{colim} gives $\Delta g/g \approx 0.01$. This estimate follows from using  $g(z)=g_0 e^{-z/\lambda_{\text{ev}}}$ and $z(t)=z_0+v_\perp (t-T/2)$, where $z$ is the distance of the atom from the cavity surface, and $v_{\perp}$ is the component of the atom's velocity perpendicular to the cavity surface. Expanding $g(z(t))$ to first order in $v_\perp(t-T/2)/\lambda_{\text{ev}}$ gives us $g(t)=g-[g v_\perp T/\lambda_{\text{ev}}](t-T/2)/T$, where $g=g_0 e^{-z_0/\lambda_{\text{ev}}}$. Finally, we have $\Delta g/g=v_\perp T/\lambda_{\text{ev}}$ with $\lambda_{\text{ev}} \approx 1000$ nm.
{\color{black} We now consider the Doppler shift. Since the proposed atomic levels have different values of $e$--$g$ and $e$--$s$ frequencies, the value $\Delta_{p1}$ of the Doppler shift to $\Delta_1$ will be different from the value $\Delta_{p2}$ of the Doppler shift to $\Delta_{2}$. As discussed previously in Sec.~\ref{atomiclsch}, we can use a chiral cavity (with the one relevant optical mode), ensuring we do not have two optical modes coupled to the same transition with opposite Doppler shifts. As shown in Fig.~\ref{cavsk}, using the control pulse at an angle to the cavity, we can ensure both Doppler shifts are equal i.e.,~$\Delta_{p1} = \Delta_{p2}=\Delta_d$, where $\Delta_d$ denotes the common value of both the Doppler shifts.  The scheme shown in Fig.~\ref{cavsk} works for both lambda and ladder atomic structures, provided one suitably chooses the propagation direction of the control pulse relative to the cavity. Another method is to measure the speed of the atom using classical light
and adjust the frequencies of the control pulse and the cavity fields, eliminating both Doppler shifts. The cavity resonance frequency can be varied by changing the temperature of the cavity \cite{cavityres} or using electro-optic modulation \cite{cavityresln}. Then, assuming $\Delta_{p1}=\Delta_{p2}=\Delta_d$, we find that, for single-photon retrieval, the contribution to the infidelity for case 1 parameters in Table \ref{tab1} scales as $a_2(\Delta_d/g)^2$ for $\abs{\Delta_d/g}<0.25$. For the example shown in Fig.~\ref{dopplererr}(a), we have $  a_2 =   2.5$. For case 2 parameters in Table \ref{tab2}, the contribution to the infidelity scales as $d_1 (\Delta_d/g)+d_2 (\Delta_d/g)^2$ for $\abs{\Delta_d/g}<0.25$ with $d_1 \ll d_2$. For the example shown in Fig.~\ref{dopplererr}(b), we have $(d_1, d_2) = (0.055, 1.16)$. In cases 1 and 2, Doppler-induced inefficiencies for single-photon absorption obey similar error scalings as the corresponding inefficiencies for single-photon generation. For atoms at room-temperature speeds and the proposed atomic level scheme, $\Delta_d \approx (2\pi)0.38$ GHz, which corresponds to $\abs{\Delta_d/g} \approx 0.24$ for the model cavity parameters considered in Fig.~\ref{dopplererr}.

\begin{figure}[tp]
            \includegraphics[
		width=\columnwidth
    		]{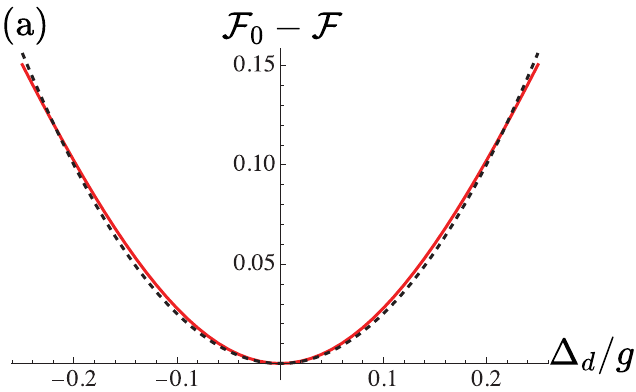}
            \includegraphics[
		width=\columnwidth
    		]{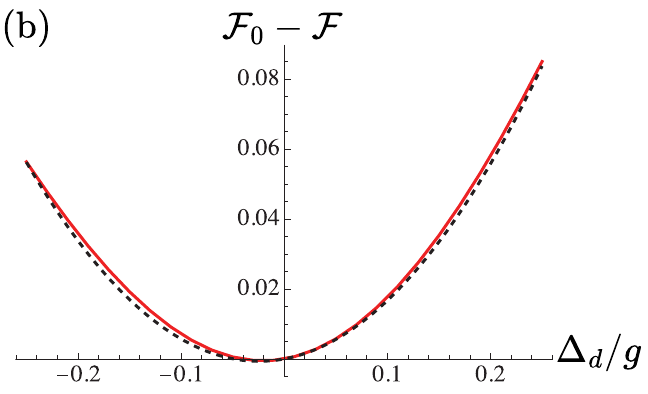}
	\caption{ We plot here the correction to the fidelity of the retrieved photon $\mathcal{F}_0-\mathcal{F}$ due to Doppler-shifted laser and cavity frequencies. We assume both the control laser and cavity field have the same shift $\Delta_d$ (as discussed in Sec.~\ref{expgf}). (a) The red line corresponds to parameters in model cavity 1a in Table \ref{tab1} (for case 1). The correction is described, approximately, by the fit $2.5 (\Delta_d/g)^2$ (shown by the black-dashed line). (b) The red line corresponds to parameters in model cavity 5 in Table \ref{tab2} (for case 2). The correction is described, approximately, by the fit $0.055  (\Delta_d/g) + 1.16 (\Delta_d/g)^2$ (shown by the black-dashed line). }
	\label{dopplererr}
\end{figure}

\section{Atom-photon controlled-$Z$ gate} \label{cpfsec}

We now outline the physics of the atom-photon controlled-$Z$ gate. We first consider the simplest case where a single chiral mode $a$ interacts with the atom. This gate has been demonstrated using a trapped atom inside a Fabry-Perot cavity \cite{atmph, atmph2} and has several uses: cluster state generation (as discussed in Appendix~\ref{clustersec}), quantum communication (as discussed in Appendix~\ref{comsec}), and, as shown in Fig.~{\ref{pcpfdet}}, non-destructive measurement of photons \cite{qnd}. Furthermore, as shown in Fig.~\ref{ppcpf}, this gate can be used to implement the photon-photon controlled-$Z$ gate \cite{Duan}. 

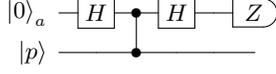
\begin{figure}[tp]
\vspace{0.5cm}
\hspace{1cm}
\Qcircuit @C=0.8em @R=0.9em {
\lstick{\ket{0}_a}& \gate{H} & \ctrl{1}  & \gate{H} & \qw & \measureD{Z}  \\
\lstick{\ket{p}}&  \qw & \control \qw & \qw & \qw & \qw \\
} 
\caption{Protocol for using the atom-photon controlled-phase gate (represented by two black circles connecting the atom and the photon) to implement non-destructive detection of a single photon \cite{qnd}.  If there is no incoming photon ($\ket{p}=\ket{0}$), the atomic state is unchanged. Otherwise, when there is an incoming photon ($\ket{p} = \ket{1}$), the atomic state flips to $\ket{1}_a$. }
\label{pcpfdet}
\end{figure}

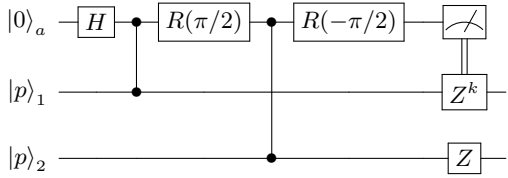
\begin{figure}[t]
\vspace{0.5cm}
\hspace{4cm}
\Qcircuit @C=0.8em @R=1.4em {
\lstick{\ket{0}_a}& \gate{H} & \ctrl{1}  & \gate{R(\pi/2)} & \ctrl{2} & \gate{R(-\pi/2)}&  \qw & \meter \cwx[1] \\
\lstick{\ket{p}_1}&  \qw & \control \qw & \qw & \qw & \qw & \qw& \gate{Z^k} & \qw \\
\lstick{\ket{p}_{2}}&  \qw & \qw & \qw & \control \qw & \qw & \qw& \gate{Z} & \qw\\
}
\caption{Protocol for the photon-photon controlled-phase gate \cite{Duan}. The black circles connecting the atom and a photon represent the atom-photon controlled-phase gate. The unitary $R(\theta)$ is defined by $R(\theta)\ket{0}=\cos(\theta/2)\ket{0}+\sin(\theta/2)\ket{1}$ and $R(\theta)\ket{1}=-\sin(\theta/2)\ket{0}+\cos(\theta/2)\ket{1}$. The correction unitary on the first photon depends on the atom's measurement outcome $k$  in the $Z$ basis. This scheme works for both single-rail photons (encoded in the Fock basis) and dual-rail photons (encoded in the polarization basis). For single-rail photons, $\ket{p} \in \{\ket{0}, \ket{1}\}$, and for dual-rail photons, $\ket{p} \in \{\ket{v}, \ket{h}\}$.}
\label{ppcpf}
\end{figure}

We have the cavity-atom Hamiltonian in the rotating frame (with the frequency of the input light $\tilde{\omega}_p$) given by
\begin{align}
H&= \da \ketbra{e}{e} + \dc a^\dagger a + (g a^\dagger \ketbra{g}{e} + \text{h.c.~}  ),\\
\da&=\omega_{eg}-\tilde{\omega}_p, \notag\\
\dc&=\omega_{c}-\tilde{\omega}_p, \notag
\end{align}
where $a$ is the cavity-mode annihilation operator, $\omega_c$ is the cavity resonance frequency, $\tilde{\omega}_p$ is the frequency of the incoming pulse, and $g$ is the single-photon Rabi frequency. The following equations \cite{kuzmich} 
\begin{align}
\dot{a}&=-i[a, H] -\kappa a -\sqrt{2\kappa_{\text{ex}}}a_{\text{in}},\label{in1} \\a_{\text{out}}&=a_{\text{in}}+\sqrt{2\kappa_{\text{ex}}}a \label{in2}
\end{align}
can be used to solve for the transmission coefficient \cite{parkins}. For input photons resonant with the cavity and the atom ($\da = \dc =0$), in the limit $\kappa_{\text{ex}}T \rightarrow \infty$, we obtain the following transmission coefficient $ t=a_{\text{out}}/a_{\text{in}}$ (for details of the derivation, see Appendix \ref{cpf} and Eq.~(\ref{tcf}))
\begin{align}
t=\frac{C-1+\kappa_{\text{i}}/\kappa_{\text{ex}}}{C+1+\kappa_{\text{i}}/\kappa_{\text{ex}}}, \label{tcoeff}
\end{align}
where $C=\frac{ g^2}{\kappa \gamma}$ is the cooperativity. Here $\abs{t}$ is always less than one due to atomic coherence decay rate $\gamma$ (when the atom is coupled, i.e.\ $g \neq 0$) and the cavity's intrinsic loss rate $\kappa_{\text{i}}$.

We now consider the case where our input photons (resonant with the bare cavity) have two orthogonal polarizations, $h$ and $v$, and one of these, say $v$, does not couple to the cavity \cite{Duan, nonlinearvolz}. Using a third atomic level $\ket{s}$, which does not couple to the cavity, we get a conditional phase gate acting on the atom-photon state i.e.,~$\hat{Z}_{a p}=\ketbra{g}{g}_a \otimes I_p + \ketbra{s}{s}_a\otimes (\ketbra{v}{v}_p-\ketbra{h}{h}_p)$. For both $\ket{g}$ and $\ket{s}$ atomic states, $v$-polarized light (uncoupled from the cavity) passes through the waveguide without interacting with the cavity; therefore its phase remains unchanged. For the case when the atom is in state $\ket{g}$, the atom-cavity eigenstates are shifted by $\pm g$, enabling $h$-polarized light (resonant with the bare cavity) in the waveguide to pass through without interacting with the cavity.  In the case where the atom is in the state $\ket{s}$, the $h$-polarized light, now resonant with the cavity, passes through the cavity obtaining a phase shift of $\pi$. For single-rail photonic qubits encoded in mode $h$, we can write the gate as $\hat{Z}_{ap}=\ketbra{g}{g}_a \otimes I_p + \ketbra{s}{s}_a \otimes (\ketbra{0}{0}_p-\ketbra{1}{1}_p)$ (Fig.~\ref{pcpfdet} shows how to implement photon detection using this gate). As in the case of single-photon retrieval, this gate can be passively multiplexed (as shown in Fig.~\ref{MulCh}).

We now consider the non-chiral cavity. In this system, the transmission coefficient calculation gives us the gate $\tilde{Z}_{ap}$ defined as follows:
\begin{align}
 \tilde{Z}_{ap} =& \ketbra{g}{g}\otimes \Bigb{e^{-i2\phi}\ketbra{R_h}{L_h}+e^{i2\phi}\ketbra{L_h}{R_h}}  \label{chiralCPF} \\
&-\ketbra{s}{s}\otimes \Bigb{\ketbra{R_h}{R_h}+\ketbra{L_h}{L_h}},\notag
\end{align}
where $\ket{R_h(L_h)}$ denotes a right-moving (left-moving) photon with polarization $h$ that couples to the atom-cavity system.
As shown in Appendix \ref{nonchiralcal}, the controlled-phase gate in the non-chiral case can be implemented using two different schemes. In the first scheme, the phase in $g=|g| e^{i\phi}$ is measured by interferometry using weak classical pulses and the atom-photon interaction from Eq.~(\ref{chiralCPF}). 
Using this information, a gate sequence that depends on $\phi$ can be applied to implement the atom-photon controlled-phase gate (up to an overall minus sign) in the basis $\{\ket{sL_h}, \ket{sR_h}, \ket{gL_h} , \ket{gR_h} \}$ [see Appendix \ref{cpf} more details]. 
This version of the protocol can be actively multiplexed (as shown in Fig.~\ref{MulChNC}) i.e.,~the atom's position in the active cavity and the cavity phase $\phi$ are monitored. In the second scheme, the phase measurement is not needed. In this scheme, the following gate sequence can be used to implement the controlled-phase gate in the basis $\{\ket{g R}_{v}, \ket{g R}_{h},\ket{s R}_{v} \ket{s R}_h \}$:
\begin{align}
 \tilde{Z}_{ap}^{\prime}U_p \tilde{Z}_{ap}^{\prime} =& \ketbra{g}{g}\otimes \Bigb{\ketbra{R}{R}_{v}+\ketbra{R}{R}_{h}} \label{nomeasurementgate}\\&+ \ketbra{s}{s}\otimes\Bigb{\ketbra{R}{R}_v - \ketbra{R}{R}_h} \notag \\&+ \ketbra{g}{g}\otimes \Bigb{\ketbra{L}{L}_v - \ketbra{L}{L}_h} \notag \\&+ \ketbra{s}{s}\otimes \Bigb{\ketbra{L}{L}_v + \ketbra{L}{L}_h}. \notag
\end{align}
Here the $h$ polarization couples to the atom-cavity system, while the $v$ polarization is uncoupled, $\tilde{Z}_{ap}^{\prime}$, defined in Eq.~(\ref{zapt}), is the non-chiral-cavity-based atom-photon controlled-phase gate in the basis $\{\ket{R}_v, \ket{R}_h, \ket{L}_v, \ket{L}_h \}$, and $U_p=-\ketbra{R}{R}_{h}+\ketbra{L}{L}_{h} + \ketbra{R}{R}_v+\ketbra{L}{L}_v$ (implemented electro-optically using non-reciprocal effects \cite{electro_nonreci}). The main idea is that we can send our photon to the system twice (hence two applications of $\tilde{Z}_{ap}^{\prime}$) to get rid of the unwanted phase $e^{i2\phi}$. However, this must be done in a gate sequence that keeps the desired conditional $-1$ phase intact. The main technical constraint in using this protocol is ensuring that the single-photon duration $T$ is much smaller than the time it takes for the phase of the cavity field to change. This protocol can be passively multiplexed, as shown in Fig.~\ref{MulCh}, since the phase of the cavity field at the position of the atom does not need to be monitored. 
For more details, see discussion in Appendix \ref{cpf}.

\subsection{Experimental considerations} \label{cpfconsider}

We now consider how non-idealities in the experiment affect the performance of the controlled-phase gate. For simplicity, we consider the chiral version of the gate. We first consider the effect of finite pulse duration on the fidelity of the output state for the chiral case. In the case $\dc = \da = 0$ and in the adiabatic limit ($\kappa_{\text{ex}}T \gg 1$),
 we find the following
 state fidelities (see Eqs.~(\ref{nonadF}) for details), where $F=\abs{\int dt a_{\text{out}}a_{\text{in}}^*}^2$, for the case where the cavity is empty ($F_1$), and where the cavity has an atom ($F_2$):
\begin{subequations}
\begin{align}
\sqrt{F_1}&=1-\Bigbl{\frac{2\kappa_{\text{i}}}{\kappa}+\frac{8\pi^2}{3}\Bigb{1-\frac{\kappa_{\text{i}}}{\kappa}}\Bigb{\frac{1}{\kappa T}}^2}, \label{fid1} \\
\sqrt{F_2}&=1- \frac{\kx}{\kappa}\Bigbl{C^{-1}+\frac{8\pi^2}{3}\Bigb{\frac{\kappa}{g}}^2 \Bigb{\frac{1}{gT}}^2}. \label{fid2}
\end{align}
\end{subequations}
Here, the terms proportional to $1/T^2$ are the leading error terms corresponding to the finite duration of the input light. When the cavity is empty, light enters the cavity and suffers photon loss due to the intrinsic loss $\kappa_{\text i}$. When the atom is coupled to the cavity, light is lost due to decay $\gamma$ described by the term proportional to $C^{-1}$. We now briefly comment on the effect of the variation of the coupling strength $g$ in time by an amount $\Delta g$ (see Sec.~\ref{expgf} for the precise definition of $\Delta g$). The lowest-order (in $\Delta g/g$) correction to the state fidelity is third-order in $\epsilon_i$, where  $\epsilon_1=\gamma/g$, $\epsilon_2=\dc/g$, and $\epsilon_3=\da/g$. Since $\epsilon_i \ll 1$, the error is negligible. For details, see Appendix~{\ref{cpf}} and Eq.~(\ref{gfid}).

We now compare the fidelity of our controlled-phase gate with the fidelities of entangling gates in other platforms.
The fidelity of the entangled state is defined as $\mathcal{F}_{\text{en}}=\bra{\Phi^{+}_{ap}}\rho_{ap} \ket{\Phi^{+}_{ap}}$, where $\rho_{ap}$ is the output state obtained by applying the controlled-phase gate on the initial state $\ket{+}_a\ket{+}_p$. Here subscripts $a$ and $p$ denote the atom and photon, respectively.  
The ideal state is $\ket{\Phi^+_{ap}}\propto \ket{0}_a \ket{+}_{p}+\ket{1}_a \ket{-}_{p}$. The values of $\mathcal{F}_{\text{en}}$ in Table \ref{tab1} compare well with $\mathcal{F}_{\text{en}}=0.944$ obtained in the state-of-the-art spin-photon system based on a single silicon-vacancy color center integrated inside a diamond nanophotonic cavity \cite{qkdlk}.

\section{APPLICATIONS}\label{apps}
Following work in Refs.~\cite{Lind, pich}, we can use the coherent emission process, atomic state rotations, and the controlled-phase gate to make protocols for generating cluster states. The protocols are outlined in Appendix~\ref{clustersec}. Moreover, the atom-photon interaction can be used to implement a quantum communication protocol that enables two users to set a shared secret key \cite{qkdlk}. Complete protocols for this task using photons encoded in single-rail and dual-rail bases are presented in Appendix \ref{comsec}.

\begin{acknowledgements}
We thank Kunal Sharma, Paul Lett, Matthew Hummon, Xiyuan Lu, and Roy Zektzer for helpful discussions. S.A., D.D., and A.V.G.~were supported in part by DARPA SAVaNT ADVENT, the DoE ASCR Quantum Testbed Pathfinder program (awards No.~DE-SC0019040 and No.~DE-SC0024220), NSF QLCI (award No.~OMA-2120757), NSF STAQ program, AFOSR MURI,   ARL (W911NF-24-2-0107), and NQVL:QSTD:Pilot:FTL. S.A., D.D., and A.V.G.~also acknowledge support from the U.S.~Department of Energy, Office of Science, National Quantum Information Science Research Centers, Quantum Systems Accelerator (QSA), and from the U.S.~Department of Energy, Office of Science, Accelerated Research in Quantum Computing, Fundamental Algorithmic Research toward Quantum Utility (FAR-Qu).  K.H., F.Z., and K.S. were supported in part by the DARPA SAVaNT program and the Army Research Office under Cooperative Agreement Number W911NF-21-2-0106. D.D.~acknowledges support by the NSF GRFP under Grant No.~DGE-1840340 and an LPS Quantum Graduate Fellowship.
\end{acknowledgements}

\bibliographystyle{apsrev4-2}
%

\appendix

\section{Derivation of the control pulse shapes for single-photon retrieval and absorption} \label{pulseshape}

In this section, we continue our discussion from Sec.~\ref{singlephoton} and provide details on how we compute the  control pulse shape $\Omega_0(t)e^{i\phi_0(t)}$ to retrieve/absorb a single-photon pulse with mode shape $h(t)$ and frequency $\omega_c +\Delta_p$ with efficiency $\eta_{\text{r}}$. For the chiral case, we can always choose $g$ to be real by redefining mode $a$. For the non-chiral case, we can absorb the phases into the redefined single mode, $ (e^{i\phi}a + e^{-i\phi} b)/\sqrt{2} \rightarrow a $, and then take $g$ in the atom-cavity Hamiltonian to be real. For our convenience, we adjust the phase in the Hamiltonian in Eq.~(\ref{Hcoherent}) using $\ket{e}\rightarrow i\ket{e}$ to obtain the equations
\begin{align}
i\dot{c}_s&=(\Delta_1 - \Delta_2)c_s-i\Omega^* c_e, \label{eom1}\\
i\dot{c}_e&=i\Omega c_s -(\Delta_2+i\gamma)c_e+igc_g, \label{eom2}\\
i\dot{c}_g&=-ig c_e -i\kappa c_g, \label{eom3}
\end{align}
where $\Delta_1=\omega_1 -\omega_{es}$ and $\Delta_2 = \omega_c-\omega_{eg} $. Here $c_s$, $c_e$, and $c_g$ are the amplitudes of states $\ket{s, 0}$, $\ket{e, 0}$, and $a^\dagger \ket{g, 0}$, respectively. 
Following Ref.~\cite{gauss}, we can set $a_{\text{out}}=\sqrt{2\kx}c_g=\sqrt{\eta_{\text{r}}}h(t)e^{-i\Delta_pt}$, where $a_{\text{out}}$ is the output single-photon wavefunction, to solve for the necessary $\Omega_0(t)$ and $\phi_0(t)$ as follows:
\begin{align}
\Omega_0(t)&=\frac{\abs{z}}{\sqrt{\rho_{ss}}}, \label{controlP}\\
\phi_0(t)&=(\Delta_1 - \Delta_2)t + \arg{(z)}+i\int_0^t dt^\prime \> \frac{f(t^\prime)}{2\rho_{ss}(t^\prime)}, \label{controlphi}
\end{align}
where
\begin{align}
z&= \dot{c}_e +(\gamma-i\Delta_2)c_e -gc_g,\\
f&=zc_e^* -z^* c_e,\\
\dot{\rho}_{ss}&=-(z c_e^* + c_e z^*),
\end{align}
where $\rho_{ss} = \lvert c_{s} \rvert^2$. Note here that $f$ and $\dot{\rho}_{ss}$ are slowly varying functions of time. We note that $z$, $f$, and $\dot{\rho}_{ss}$ can be written in terms of $c_g=a_{\text{out}}/\sqrt{2\kx}$ using Eqs.~(\ref{eom1}--\ref{eom3}) [$c_e$ can be written in terms of $c_g$ and $\dot{c}_g$ using Eq.~(\ref{eom3})]. We also note that $\arg(z) \approx -\Delta_p t+\text{const.}$, which means that setting $\Delta_1=\Delta_2+\Delta_p$ ensures that the additional correction to the control pulse's phase is a slowly varying function. In the adiabatic limit ($\kappa_{\text{ex}} T \gg 1$), $\Omega_0(t)$ and $\phi_0$ can be computed by writing $z$, $f$, and $\rho_{ss}$ in terms of $a_{\text{out}}$ and approximating $\dot h$ as zero to obtain
\begin{align}
\Omega_0(t) &= \frac{\abs{\xi_x+i\xi_y}}{g}\frac{A(t)}{[1-(\eta_{\text{r}}/\eta_{\text{max}})\int_0^t h^2 dt^\prime]^{1/2}},\\
\phi_0(t)=&\pi + \text{arg}(\xi_x + i\xi_y) \label{phiexp1}\\
           &+ \eta_{\text{max}}\frac{(\kappa \xi_y+\Delta_p \xi_x)}{2g^2 \kx} \ln(1-(\eta_{\text{r}}/\eta_{\text{max}}) \int_0^t dt^\prime h^2), \notag \\
\xi_x=& g^2 -\Delta_p^2 -\Delta_2 \Delta_p + \kappa \gamma,\\
\xi_y=& -\gamma \Delta_p - \kappa(\Delta_p +\Delta_2),\label{phiexp3}
\end{align}
where $A(t)=(\eta_{\text{r}}/ 2\kx)^{1/2}h(t)$.
For single-photon absorption, we use the time-reversed version of the control pulse $\Omega_0(t)\rightarrow -\Omega_0(T-t)$ and $\phi_0(t)\rightarrow -\phi_0(T-t)$ (see discussion in Sec.~\ref{singlephoton} for details).

For an atomic level scheme such that the $s$ state has higher energy than the $e$ state, and the $e$ state has the largest decay rate, we can use the same control pulse as above, except that we redefine $\Delta_1$ to $\omega_{se}-\omega_1$, where $\omega_{se}$ is the energy of the $s$--$e$ transition. 
\section{Other variations of cavities and their phase shift calculation} \label{cpf}

In this section, we present the details of the phase shift calculation (presented in Sec.~\ref{cpfsec}) for both chiral and non-chiral microcavities. We follow Ref.~\cite{parkins}.

\subsection*{Case 1: Chiral setup}\label{awd}

We compute here the transmission coefficient in the chiral mode~case discussed in Sec.~\ref{cpfsec}. We excite the $a$ mode by sending in light through mode $a_{\text{in}}$. The cavity-atom Hamiltonian in the rotating frame is 
\begin{align}
H&= \da \ketbra{e}{e} + \dc (a^\dagger a + b^\dagger b) + h(a^\dagger b + b^\dagger a ) \notag\\
& \hspace{1.5 cm} + (g a^\dagger \ketbra{g}{e} + g a\ketbra{e}{g} ), \label{ChHam}\\
\da&=\omega_{eg}-\tilde{\omega}_p,\\
\dc&=\omega_{c}-\tilde{\omega}_p,
\end{align}
where $\tilde{\omega}_p$ is the frequency of the incoming photon pulse, and $h$ is the coupling between the two modes due to backscattering.
The equations of motion are
\begin{align}
\dot{c}_{g, a}&= -(\kappa + i\dc)c_{g, a} - ih c_{g, b} -i g c_{e} \label{eomc1}\\ &-\sqrt{2\kappa_{\text{ex}}} a_{\text{in}}(t) , \notag \\
 \dot{c}_{g, b}&=-(\kappa + i\dc)c_{g, b} - ih c_{g, a},  \\
\dot{c}_{e}&=-(\gamma+i\da)c_{e} - i g c_{g, a} \label{eomc3},
\end{align}
where $c_{g, a}$, $c_{g, b}$, and $c_{e}$ are the amplitudes of states $a^\dagger\ket{g, 0}$, $b^{\dagger}\ket{g, 0}$, and $\ket{e, 0}$, respectively, and $c_{a(b), \text{in}}$ is the single-photon input to mode $a(b)$ (as shown in Fig.~\ref{mt1}). We can then use the Fourier transform
\begin{align}
\tilde{A}(\omega)&=\frac{1}{\sqrt{2\pi}}\int dt\> e^{i\omega t}A(t),\\
A(t)&=\frac{1}{\sqrt{2\pi}}\int d\omega\> e^{-i\omega t}\tilde{A}(\omega)
\end{align}
to obtain a set of linear equations. For $h=0$, we have the following transmission coefficient $t$:
\begin{align}
 t=\frac{ a_{\text{out}}}{ a_{\text{in}}}= \frac{\kappa_{\text{i}} +i \Delta_{\text{cp}} + \frac{g^2}{i\Delta_{\text{ap}}+\gamma} - \kappa_{\text{ex}}}{\kappa_{\text{i}}  + i\Delta_{\text{cp}}+ \frac{g^2}{i\Delta_{\text{ap}}+\gamma} + \kappa_{\text{ex}}}. \label{tcf}
\end{align}
On resonance (i.e.,\ when $\Delta_{\text{cp}}= \Delta_{\text{ap}}=0$), we get Eq.~(\ref{tcoeff}), where $C=\frac{ g^2}{\kappa \gamma}$. The leading error terms in the state fidelity $F=\abs{\int dt a_{\text{out}}a_{\text{in}}^*}^2$ can be computed from solving Eqs.~(\ref{eomc1}--\ref{eomc3}) in the adiabatic limit ($\kappa_{\text{ex}}T \gg 1$). For $\Delta_{\text{cp}}=\Delta_{\text{ap}}=0$ and $h=0$, we have
\begin{subequations} \label{nonadF}
\begin{align}
1-\sqrt{F_1}&=\frac{2\kappa_{\text{i}}}{\kappa}+\frac{8\pi^2}{3}\Bigb{1-\frac{\kappa_{\text{i}}}{\kappa}}\Bigb{\frac{1}{\kappa T}}^2, \\
1-\sqrt{F_2}&=\frac{\kx}{\kappa}\Bigbl{C^{-1}+\frac{8\pi^2}{3}\Bigb{\frac{\kappa}{g}}^2 \Bigb{\frac{1}{gT}}^2},
\end{align}
\end{subequations}
where $F_1 (F_2)$ is the state fidelity wherein the atom is uncoupled from (coupled to) the cavity. We use $a_{\text{in}} \propto \sin^2(\pi t/T)$.

We now consider the case where $g(t) = g_0 - \Delta g f(t)$, where $f$ is a slowly varying function that goes from 0 to 1. The amplitude $c_g$ can be solved using the ansatz $c_g =\sum_{n=0} c_{gn}\beta^n$, where $\beta=\Delta g/g_0$, giving the equation
\begin{align}
\ddot{c}_{gn}&=-\alpha c_{gn}-\sigma \dot{c}_{gn}+W_n, \label{gfid}
\end{align}
where 
\begin{align}
\alpha&=g_0^2 + (\gamma + i\da)(\kappa + i\dc),\\
\sigma&=(\kappa + \gamma)+i(\da + \dc),\\
W_0&=-\sqrt{2\kx}(\gamma + i\da)a_{\text{in}}-\sqrt{2\kx}\dot{c}_{\text{in}}\\
W_1&=-\sqrt{2\kx}a_{\text{in}} \dot{f}-\dot{c}_{g0}\dot{f} \notag \\
&\>\>\>\>\>\>+c_{g0}[2g_0^2 f -\dot{f}(\kappa+i\dc)].
\end{align}
We note here that $c_{g1}$ is coupled to $c_{g0}$ via $W_1$.
The equations can be solved using Fourier transformation. The lowest-order (in $\Delta g/g_0$) correction to the state fidelity is third-order in $\epsilon_i$, where  $\epsilon_1=\gamma/g_0$, $\epsilon_2=\dc/g_0$, and $\epsilon_3=\da/g_0$. Since $\epsilon_i \ll 1$, the error is negligible.

\subsection*{Case 2: Non-chiral setup} \label{nonchiralcal}

Here, we study the atom-photon controlled-phase gate for the non-chiral case, wherein both modes $a$ and $b$ couple to the atom-cavity system. The  atom-cavity  Hamiltonian in the rotating frame is
\begin{align}
H&= \da \ketbra{e}{e} + \dc (a^\dagger a + b^\dagger b) + h(a^\dagger b + b^\dagger a ) \notag\\
 &+ g(  e^{-i\phi} a^\dagger \ketbra{g}{e} +  e^{i\phi} a \ketbra{e}{g} ) \notag \\&+ g( e^{i\phi} b^\dagger \ketbra{g}{e} +  e^{-i\phi} b\ketbra{e}{g} ), \label{HamTE}\\
\da&=\omega_{eg}-\tilde{\omega}_p,\\
\dc&=\omega_{c}-\tilde{\omega}_p.
\end{align}
We note here that the above Hamiltonian can be mapped to the Hamiltonian from the chiral case in Eq.~(\ref{ChHam}) by mapping $(e^{i \phi} a + e^{i \phi} b)/\sqrt{2} \rightarrow a$ and $g \sqrt{2} \rightarrow g$
(and mapping $(e^{i \phi} a_{\text{in}} + e^{i \phi} b_{\text{in}})/\sqrt{2} \rightarrow a_{\text{in}}$ and $(e^{i \phi} a_{\text{out}} + e^{i \phi} b_{\text{out}})/\sqrt{2} \rightarrow a_{\text{out}}$ accordingly).
However, in contrast to the chiral case in which we drive the mode that couples to the atom, here we choose to separately drive modes $a$ and $b$ which only partially couple to the mode $(e^{i \phi} a + e^{i \phi} b)/\sqrt{2}$.
The equations of motion can then be written as
\begin{align}
\dot{c}_{g, a}&= -(\kappa + i\dc) c_{g,a}  - ih  c_{g,b}  -i ge^{-i\phi}  c_{e} -\sqrt{2\kappa_{\text{ex}}} a_{\text{in}}(t) ,   \\
  \dot{c}_{g, b} &=-(\kappa + i\dc) c_{g, b}   - ih  c_{g, a}   -i g e^{i\phi} c_{e} -\sqrt{2\kx}b_{\text{in}}(t) , \label{bbin} \\
\dot{c}_{e}&=-(\gamma+i\da)  c_{e} - i g e^{i\phi}c_{g,a} -i ge^{-i\phi}   c_{g, b}, \label{l1}
\end{align}
where $c_{g, a}$, $c_{g, b}$, and $c_{e}$ are the amplitudes in the states $a^\dagger\ket{g, 0}$, $b^{\dagger}\ket{g, 0}$, and $\ket{e, 0}$, respectively, and $a_{\text{in}}(b_{\text{in}})$ is the single-photon input to the mode $a(b)$ (see Fig.~\ref{mt1}).
Then, for $\Delta_{\text{ap}}=\Delta_{\text{cp}}=0$, $\kappa_{\text{i}}\ll \kx$, and $h=0$, we have
\begin{align}
 c_{g, a}&= -\frac{\sqrt{2\kx}}{\kappa}
 \Bigbl{\frac{(1 + C) a_{ \text{in}}  - C e^{-i2\phi} b_{\text{in}}}{1+2C}},\\
 c_{g, b}&=- \frac{\sqrt{2\kx}}{\kappa}\Bigbl{\frac{(1 + C) b_{\text{in}} -  C e^{i2\phi} a_{\text{in}}}{1+2C}},
\end{align}
where $\phi$ is defined in Eq.~(\ref{HamTE}), and $C=g^2/\kappa \gamma$. We expect nonzero values of $h$ that satisfy $h \ll \kx$ to reduce the fidelity of the states only marginally.
When the atom is not coupled to the cavity ($g=0$) and $\kappa_{\text{i}} \ll \kx$, we have
\begin{subequations}
\begin{align}
  a_{\text{out}} &=-  a_{\text{in}},\\
  b_{\text{out}}&=-  b_{\text{in}}.
\end{align}
\end{subequations}
For strong atom-cavity coupling ($C \gg 1$) and $\kappa_{\text{i}} \ll \kx$, we get
\begin{subequations} \label{phs2}
\begin{align}
 a_{\text{out}} &=e^{-i2\phi}   b_{\text{in}},\\
 b_{\text{out}} &=e^{i2\phi}   a_{ \text{in}}.
\end{align}
\end{subequations}

We can then write this process as the unitary
\begin{align}
 \tilde{Z}_{ap} =& \ketbra{g}{g}\otimes \Bigb{e^{-i2\phi}\ketbra{R}{L}+e^{i2\phi}\ketbra{L}{R}}   \notag\\
&-\ketbra{s}{s}\otimes \Bigb{\ketbra{R}{R}+\ketbra{L}{L}}, \label{rechiralCPF}
\end{align}
where $\ket{R(L)}$ corresponds to the right-moving (left-moving) single photon leaking to the waveguide through the cavity mode $a(b)$ (see also Fig.~\ref{mt1}).

Now, this gate can be used to implement the atom-photon controlled-phase gate within the photon basis $\{\ket{R}, \ket{L} \}$ by using the sequence $H_p V  \tilde{Z}_{ap}V^{\dagger} H_p$, where $V=e^{-i2\phi} \ketbra{L}{L}+\ketbra{R}{R}$ (implemented using electro-optics), and $H_p$ is the Hadamard gate in the $\{R, L \}$ basis. This assumes we can measure the phase $\phi$ quickly enough. We then get the gate
\begin{align}
H_p V  \tilde{Z}_{ap}V^{\dagger} H_p=&\ketbra{g}{g}\otimes \Bigb{ \ketbra{R}{R} -\ketbra{L}{L}} \notag\\
& -\ketbra{s}{s}\otimes \Bigb{\ketbra{R}{R}+\ketbra{L}{L}}, \label{measurementgate}
\end{align}
which is the controlled-phase gate in the basis $\{\ket{sL}, \ket{sR}, \ket{gL} , \ket{gR} \}$ (up to an overall minus sign).

We now consider another gate sequence to implement the atom-photon controlled-phase gate without measuring the phase $\phi$. We consider single photons with polarizations $h$ and $v$; $h$ couples to the atom-cavity system, and $v$ does not. In the larger photon basis $\{\ket{R}_v, \ket{R}_{h}, \ket{L}_v, \ket{L}_h \}$, where $\ket{R(L)}_{x}$ denotes the right-moving (left-moving) photon with polarization $x$, consider the gate sequence $ \tilde{Z}_{ap}^{\prime} U_p\hat{\tilde{Z}}_{ap}^{\prime}$, where $U_p=-\ketbra{R}{R}_{h}+\ketbra{L}{L}_{h} + \ketbra{R}{R}_v+\ketbra{L}{L}_v$, and
\begin{align}
 \tilde{Z}_{ap}^{\prime}=& \ketbra{g}{g}\otimes \Bigb{e^{-i2\phi}\ketbra{R}{L}_h+e^{i2\phi}\ketbra{L}{R}_h \notag\\&+\ketbra{R}{R}_v+\ketbra{L}{L}_v}
+\ketbra{s}{s}\otimes \Bigb{-\ketbra{R}{R}_h \notag\\
&-\ketbra{L}{L}_h+\ketbra{R}{R}_v+\ketbra{L}{L}_v}. \label{zapt}
\end{align}
We then have
\begin{align}
 \tilde{Z}_{ap}^{\prime}U_p \tilde{Z}_{ap}^{\prime} =& \ketbra{g}{g}\otimes \Bigb{\ketbra{R}{R}_{v}+\ketbra{R}{R}_{h}} \notag\\&+ \ketbra{s}{s}\otimes\Bigb{\ketbra{R}{R}_v - \ketbra{R}{R}_h} \notag \\&+ \ketbra{g}{g}\otimes \Bigb{\ketbra{L}{L}_v - \ketbra{L}{L}_h} \notag \notag \\&+ \ketbra{s}{s}\otimes \Bigb{\ketbra{L}{L}_v + \ketbra{L}{L}_h}.  \label{renomeasurementgate}
\end{align}
This gate acting on the atom-photon basis states $\{\ket{gR}_v, \ket{gR}_h,\ket{sR}_v, \ket{sR}_h\}$ gives us the atom-photon controlled-phase gate.
In Fig.~\ref{echocpf}, we show the steps used to implement the atom-photon controlled-phase gate. Consider the initial state $\ket{\psi} = c_{gR_v}\ket{gR_v}+c_{gR_h}\ket{gR_h}+c_{sR_v}\ket{sR_v}+c_{sR_h}\ket{sR_h}$. In step 1, we send the photon to the cavity to obtain the state 
\begin{align}
\tilde{Z}_{ap}^\prime \ket{\psi}=& c_{gR_v}\ket{gR_v}+e^{i2\phi}c_{gR_h}\ket{gL_h}\notag\\&+c_{sR_v}\ket{sR_v}-c_{sR_h}\ket{sR_h}.  
\end{align}
In step 2, we apply the gate $U_p$, realized electro-optically using non-reciprocal effects \cite{electro_nonreci}, to obtain the state
\begin{align}
U_p\tilde{Z}_{ap}^\prime \ket{\psi}=&c_{gR_v}\ket{gR_v}+e^{i2\phi}c_{gR_h}\ket{gL_h}\notag\\&+c_{sR_v}\ket{sR_v}+c_{sR_h}\ket{sR_h}.
\end{align}
Both the right-moving and left-moving parts of the photon are routed back to the cavity as shown in Fig.~\ref{echocpf}. 
 This step can be implemented using electro-optics. This reapplies the gate $\tilde{Z}_{ap}^\prime$ in step 3 to obtain the state
\begin{align}
\tilde{Z}_{ap}^\prime U_p\tilde{Z}_{ap}^\prime \ket{\psi}=& c_{gR_v}\ket{gR_v}+c_{gR_h}\ket{gR_h}+c_{sR_v}\ket{sR_v}\notag\\&-c_{sR_h}\ket{sR_h},
\end{align}
implementing the atom-photon controlled-phase gate.
\begin{figure}[t] 
\includegraphics[width=1.0\columnwidth]{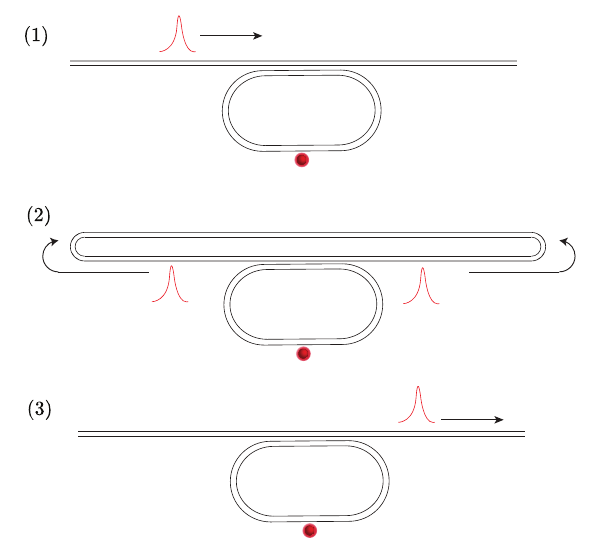}
\caption{In this figure, we show the steps used to implement the atom-photon controlled-phase gate for the non-chiral cavity without monitoring the atom.}
\label{echocpf}
\end{figure}
 Moreover, instead of using $h$ and $v$ polarizations, we can use different waveguides for dual-rail encoded photons such that one waveguide couples to the atom-cavity system and one does not.

\section{Cluster states} \label{clustersec}
 
In this section, we present the details of the protocols developed in Refs.\ \cite{Lind, edoclust} to generate cluster states using the primitives (i.e.,~single-photon retrieval/absorption and the atom-photon controlled-phase gate) studied in this work. We first consider single-rail-encoded cluster states. We relabel the atomic states considered before as $\ket{0}_a = \ket{g}$ and $\ket{1}_a = \ket{s}$. The pulse sequence for retrieval can then be represented by the operator $P_{a, i}=\ketbra{0}{0}_a \otimes I_i + \ketbra{0}{1}_a \otimes \ketbra{1}{0}_i$, where the $i$ labels the photon state's time bin, and the operator is understood to act only on atom-photon states with the photonic state in the cavity being in the vacuum. 
We note here that, since the atom moves across the cavity and the relative phase between the coupling $g$ and the control laser $\Omega(t)$ changes, the state $\ket{0}_a (\ket{0}_i+e^{i\psi}\ket{1}_i)$ will have a phase $\psi$ dependent of the position of the atom. We can ensure that the atom-photon states have a phase that is independent of the atom position for the chiral case by matching the wave-vectors of the cavity and laser fields (similar to how we cancel the Doppler contribution to the two-photon detuning for the lambda system in Fig.~\ref{cavsk}).

For example, consider the protocol for making a GHZ (Greenberger-Horne-Zeilinger) state. Starting with the state $(\ket{0}_a+\ket{1}_a)\ket{0}_1$, applying the control pulse, $P_{a,1}$, gives the state $\ket{0}_a(\ket{0}_1+\ket{1}_1)$. Applying the Hadamard gate to the atomic state (denoted by $H_a$), followed by atom-photon controlled-phase gate $\hat{Z}_{a1}$, and then reapplying $H_a$ to the atom gives the entangled state $\ket{00}_{a1}+\ket{11}_{a1}$. To pass this entanglement to the photon states 1 and 2, we apply $P_{a,2}$ to get the state $\ket{0}_a \otimes (\ket{00}_{12}+\ket{11}_{12})$. Extending this protocol as in Fig.~\ref{fock}(a) produces the $n$-photon GHZ state. Measuring the $\ket{0}_a$ state after
the last $P_{a,i}$ confirms that the atom has not escaped the cavity during the gate sequence.
In summary, for each additional photon in the state, we need to send the photon in the preceding time-bin to the cavity for an atom-photon controlled-phase gate (with Hadamard gates on the atom before and after the interaction), and a control pulse to implement $P_{a,i}$.  

We now discuss the protocol to create cluster states defined in Ref.~\cite{Rauss}. The key identity used here is
\begin{align}
&P_{a, n+1}\hat{Z}_{a \bm{j}}H_a \ket{0}_a \ket{\phi}\ket{0}_{n+1}=\ket{0}_a \hat{Z}_{n+1\bm{j}}\ket{\phi}\ket{+}_{n+1}, \notag\\
&\hat{Z}_{a\bm{j}}=\prod_{i=1}^{n^\prime}\hat{Z}_{a j_i}, \>\> i \in \{1, \hdots,  n^\prime \},
\end{align}
where $\hat Z_{a\bm{j}}$ is a controlled-phase gate between the atom and photons indexed in $\bm{j}$, $\hat Z_{n \bm{j}}$ is a controlled-phase gate between the photon labeled $n$ and photons indexed in $\bm{j}$, $\ket{\phi}$ is an arbitrary $n$-photon state, and $j_i$  (with $i \leq n^\prime$) indexes some photon states in $\ket{\phi}$. The net effect of this step is to implement the gate $\hat{Z}_{n+1 \bm{j}}$ on the $n+1$-photon state $\ket{\phi}\ket{+}_{n+1}$. If the state $\ket{\phi}$ is a cluster state, it follows that the state after the control sequence is also one. For example, consider the initial state $\ket{0}_a\ket{+}_1 \ket{0}_2$. Applying the sequence $P_{a, 2}Z_{a 1}H_a$ gives the state $\ket{0}_a \hat{Z}_{21}\ket{++}_{12}$. This can be extended to make the $n$-photon 1D cluster state (see Fig.~\ref{fock}(b)), as well as cluster states in two and three dimensions using time delays \cite{pich, edoclust}.

\begin{figure}[t] 
\includegraphics[width=0.9\columnwidth]{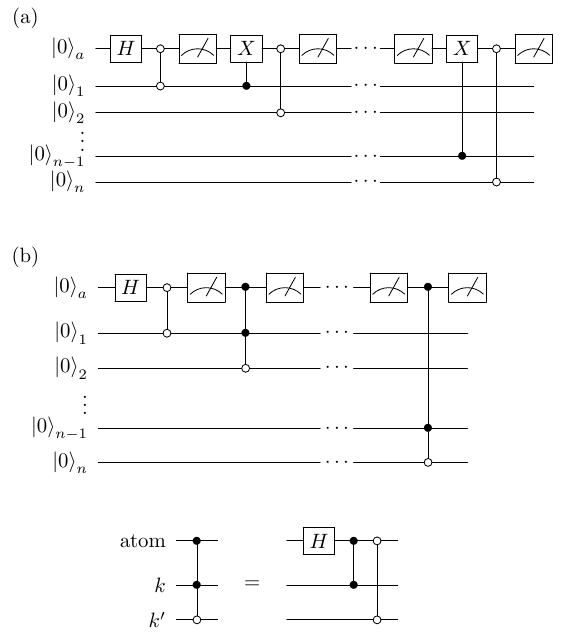}
\caption{Protocols for making (a) a GHZ state and (b) a 1D cluster state in the single-rail encoding. The two white circles that connect the atom and the state of photon $k$ represent the gate $P_{a, k}$ implemented by the control pulse. Here, the two filled circles connected by a line denote the atom-photon controlled-phase gate. The atom-photon controlled $X$ gate is implemented by sandwiching the controlled-phase gate with the Hadamard gates on the atom.}
\label{fock}
\end{figure}

Similarly, work in Refs.\ \cite{Lind, edoclust} shows that the atom-photon controlled-phase gate can be combined with atomic state rotations to create cluster states in the polarization basis. As examples, we show protocols to create GHZ states and 1D cluster states in Fig.~\ref{ghzcir} and Fig.~\ref{1dcir}, respectively. The generalization to higher dimensions is possible using time delays \cite{edoclust}.
 
\begin{figure}[t]
\includegraphics[width=0.8\columnwidth]{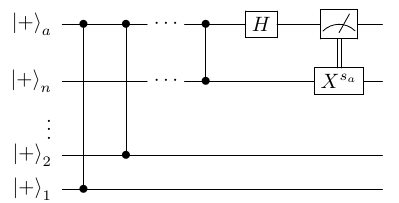}
\caption{Circuit for making the state $H^{\otimes n}\ket{\psi}$ in the polarization basis (i.e.,\ dual-rail encoding), where $\ket{\psi}$ is the $n$-photon GHZ state. Here, $s_a$ is the atomic state measurement outcome in the $Z$ basis, and the unitary after the measurement can be applied to any photon.}
\label{ghzcir}
\end{figure}

\begin{figure}
\includegraphics[width=0.9\columnwidth]{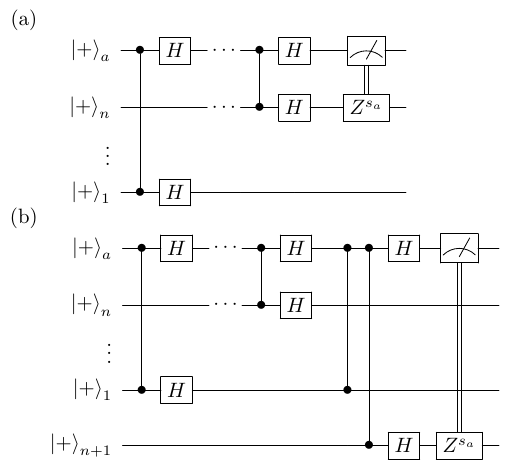}
\caption{Circuits for making 1D cluster states in the polarization basis (i.e.\ dual-rail encoding). Here $a$ labels the atomic state, and the states with numbers as subscripts correspond to photons. (a) Circuit for making 1D linear cluster states. Before measurement of the atomic state, the system's state is a 1D cluster state with the atom at one end. Measuring the atom in the $Z$ basis with outcome $s_a$ disconnects the atom. (b) Circuit for making the $(n+1)$-qubit resource state as defined by a ring graph \cite{fbqc}.} 
\label{1dcir}
\end{figure}

\section{Quantum communication} \label{comsec}

In this section, we explain how the basic primitives studied in our work (i.e.,~single-photon retrieval/absorption and the atom-photon controlled-phase gate) can be used to implement a quantum communication protocol that enables two users to set a shared secret key \cite{qkdlk}. Here, Alice and Bob wish to establish a shared secret random bit string using a central node, Charlie. Alice and Bob randomly choose the basis ($X$ or $Y$) and the state of their photonic qubits that they send to Charlie. This information is encoded in the photonic state through phase $\phi$:  $\ket{0}+e^{i\phi}\ket{1}$. The $X(Y)$ basis choice means the state will have $\phi=\{\pi, -\pi\}(\{\pi/2, -\pi/2\})$. Charlie, who has access to an atom coupled to a cavity, will perform gates and measurements on both his atomic states and the photonic states he receives from Alice and Bob.  We outline two versions of the protocol for the two possible bases: Fock states (single-rail encoding) and polarization states (dual-rail encoding), respectively. At the end of the protocol, Alice and Bob broadcast their bases, and Charlie broadcasts the $Z$-basis measurement 
results $m_i$. Instances where Alice and Bob had used different bases are discarded. Knowledge of the outcomes will allow Alice and Bob to ascertain whether their states were the same or different since the measurement outcomes, $m_i$, are distinguishable for the two cases $\phi_1+\phi_2=\{0, \pi\}$  for any basis choice, where $\phi_1$ and $\phi_2$ define Alice's and Charlie's photonic states, respectively. They can then set their shared secret key; for example, Bob can flip his bit values that disagree with Alice at the end of the protocol so that both have a secret correlated bit-string. Crucially, the measurement outcomes and the basis choice can only reveal the sum of the phases.

\subsection{Fock-basis encoding}

We now outline the communication protocol for Fock basis photons. This protocol uses control pulses (for single-photon storage), atomic state rotations, and the atom-photon controlled-phase gate. Alice sends her qubit $\ket{0}_1+e^{i\phi_1}\ket{1}_1$ to Charlie, whose atom is in state $\ket{0}_a$. Charlie applies the control laser to store the single photon so that his atomic state becomes $\ket{0}_a + e^{i\phi_1}\ket{1}_a$. Bob then sends his photon in state $\ket{0}_2+e^{i\phi_2}\ket{1}_2$ to Charlie, and Charlie applies the sequence $H_a \hat{Z}_{a2}H_a$ to obtain the state $\ket{0}_a(\ket{0}_2+e^{i\phi_1+i\phi_2}\ket{1}_2)+\ket{1}_a(e^{i\phi_1}\ket{0}_2+e^{i\phi_2}\ket{1}_2)$. Charlie then measures the atomic state in the $Z$ basis to obtain outcome $m_1$. If he measures $m_1=1$, he does nothing, and if he measures $m_1=-1$, he applies the gate $X_a$ to the atomic state, obtaining the state $\ket{0}_a (\ket{0}_2 + e^{i\phi_2+im_1 \phi_1}\ket{1}_2)$.
The next step is to store the photonic state 2 to obtain the atomic state $ \ket{0}_a + e^{i\phi_2 + im_1 \phi_1}\ket{1}_a $. Finally, Charlie applies the gate $H_a$ to his atom, measures the atom in the $Z$ basis to obtain $m_2$, and then broadcasts $m_1$ and $m_2$. See Tables \ref{tf1} and \ref{tf2} for the truth table that connects the outcomes $m_1$ and $m_2$ with $\phi_1 + \phi_2$.  

\begin{table}[htb]
\begin{tabular}{ | m{1 cm} | m{1cm}|m{1.5cm}|}
\hline
$m_1$& $m_2$ & $\phi_1 + \phi_2$  \\ \hline
 $1$ & $1$  &0 \\ \hline
 $1$ & $-1$  & $\pi$\\ \hline
 $-1$ & $1$ & 0\\ \hline
 $-1$ & $-1$  & $\pi$\\ \hline 
\end{tabular}
\caption{Truth table for the case where Alice and Bob use $x$-basis inputs. Here, $m_i$ are the $Z$-basis measurement results obtained by Charlie.}
\label{tf1}
\end{table} 
\begin{table}[htb]
\begin{tabular}{ | m{1 cm} | m{1cm}|m{1.5cm}|}
\hline
$m_1$& $m_2$ & $\phi_1 + \phi_2$  \\ \hline
 $1$ & $1$  &0 \\ \hline
 $1$ & $-1$  & $\pi$\\ \hline
 $-1$ & $1$ & $\pi$ \\ \hline
 $-1$ & $-1$  & 0 \\ \hline 
\end{tabular}
\caption{Truth table for the case where Alice and Bob use $y$-basis inputs. Here, $m_i$ are the $Z$-basis measurement results obtained by Charlie.}
\label{tf2}
\end{table}

\subsection{Polarization-basis encoding} \label{polcomm}

We now outline the communication protocol for photons encoded in the polarization basis. We use the definitions $\ket{0}=\ket{v}$ and $\ket{1} = \ket{h}$, where $h$ couples to the atom-cavity system while $v$ remains uncoupled. In the first step, Alice sends her photon in state $\ket{0}_1 + e^{i\phi_1}\ket{1}_1$ to Charlie who has his atomic qubit in state $\ket{0}_a+\ket{1}_a$. After Charlie performs the atom-photon controlled-phase gate, the atom-photon state is $\ket{0}_a(\ket{0}_1+e^{i\phi_1}\ket{1}_1) +\ket{1}_a(\ket{0}_1-e^{i\phi_1}\ket{1}_1)$. After performing the Hadamard gate on both the atom and the photon, Charlie measures the photon state in the $Z$ basis with outcome $m_1$ which results in the atomic state $\ket{0}_a+m_1e^{i\phi_1}\ket{1}_a$. We remark here that this state can also be obtained by storing the single-photon state $\ket{0}_2+e^{i\phi_1}\ket{1}_2$. In the second step, Bob sends his photonic qubit in state $\ket{0}_2+e^{i\phi_2}\ket{1}_2$ to Charlie. Charlie performs a Hadamard gate on the photon, a controlled-phase gate on the atom-photon state, and then another Hadamard gate on the photon state. This gives the atom-photon state $(\ket{0}_a+m_1e^{i(\phi_1+\phi_2)}\ket{1}_a)\ket{0}_2 + (e^{i\phi_2}\ket{0}_a + m_1 e^{i\phi_1}\ket{1}_a)\ket{1}_2$.  After measuring the photonic state in the $Z$ basis with outcome $m_2$, the atomic state is $\ket{0}_a + m_1 e^{i(\phi_1 +m_2 \phi_2)}\ket{1}_a$. After performing a Hadamard gate on the atom, Charlie measures the atomic state in the $Z$ basis to get measurement outcome $m_3$. Since $\phi_1 + \phi_2=\{0, \pi \}$, $m_3$ will have a value fixed by previous measurement outcomes. As before, the measurement outcomes depend only on the sum of the phases, $\phi_1+\phi_2$, allowing Alice and Bob to set a correlated key. See Tables \ref{t1} and \ref{t2} for the truth table that connects the outcomes with $\phi_1 + \phi_2$.
\begin{table}[htb]
\centering
\begin{tabular}{ | m{1 cm} | m{1cm}|m{1cm}|m{1.5cm}|}
\hline
$m_1$& $m_2$& $m_3$ & $\phi_1 + \phi_2$  \\ \hline
 $1$ & $1$ & $1$ &0 \\ \hline
 $1$ & $-1$ & $1$ & 0\\ \hline
 $-1$ & $1$ & $-1$ & 0\\ \hline
 $-1$ & $-1$ & $-1$ & 0\\ \hline 
 $1$ & $1$ & $-1$ & $\pi$\\ \hline
 $1$ & $-1$ & $-1$ & $\pi$ \\ \hline 
 $-1$ & $1$ & $1$  & $\pi$ \\ \hline
  $-1$ & $-1$ & $1$ & $\pi$ \\ \hline 
\end{tabular}
\caption{Truth table for $x$-basis inputs. Here, $m_i$ are the $Z$-basis measurement results obtained by Charlie.}
\label{t1}
\end{table}
\begin{table}[htb]
\centering
\begin{tabular}{ | P{1 cm} | P{1cm}|P{1cm}|P{1.5cm}|}
\hline
$m_1$& $m_2$& $m_3$ & $\phi_1 + \phi_2$  \\ \hline
 $1$ & $1$ & $1$ &0 \\ \hline
 $1$ & $-1$ & $-1$ & 0\\ \hline
 $-1$ & $1$ & $-1$ & 0\\ \hline
 $-1$ & $-1$ & $1$ & 0\\ \hline 
 $1$ & $1$ & $-1$ & $\pi$\\ \hline
 $1$ & $-1$ & $1$ & $\pi$ \\ \hline 
 $-1$ & $1$ & $1$  & $\pi$ \\ \hline
  $-1$ & $-1$ & $-1$ & $\pi$ \\ \hline
\end{tabular}
\caption{Truth table for $y$-basis inputs. Here, $m_i$ are the $Z$-basis measurement results obtained by Charlie.}
\label{t2}
\end{table}
\section{From room-temperature atoms to ultracold atoms} \label{ultracold}
In this section, we consider how various cooling methods can allow for a higher number of operations per transit event. There are several methods for slowing down atoms from the 300 m/s velocity we assume in the main text, with the resulting transit time inversely proportional to the velocity.  We list a few examples in Table \ref{tab3}, comparing the figures of merit to those obtained from using model cavity 1a in Table \ref{tab1} (using room-temperature atoms). We note that increased cooling generally comes with larger infrastructure requirements. 
The first method is Zeeman cooling, which reduces the speed of Rb atoms to 12 m/s  \cite{melentiev_zeeman_2004}, resulting in $\tau = 343$ ns for cavity 1a. The second method uses atoms released from a magneto-optical trap (MOT) which corresponds to $\tau = 1.5$ $\mu$s \cite{MOT_dayan}.
The last method uses a dipole trap which corresponds to $\tau = $ $2$ ms \cite{will_coupling_2021}. Table \ref{tab3} shows the number of operations per transit time $\tau/T$ achievable through larger values of $\tau$ using various cooling techniques.

\begin{widetext}
\begin{center}
\refstepcounter{table}
\begin{minipage}[t]{\textwidth} TABLE~\thetable. Examples of key metrics for Case~1 with different methods of cooling the atoms. Here, $g$ is the single-photon Rabi frequency, $\kappa$ is the waveguide coupling rate, $\kappa_{\text{i}}$ is the microcavity’s intrinsic loss, $2\gamma$ is the decay rate of state $e$, and $C = g^2/\kappa\gamma$ is the cooperativity. $\kappa = \kappa_{\text{ex}} + \kappa_{\text{i}}$, $T$ is the single-photon duration, and $\tau$ is the atomic transit time. Frequencies and times are in $(2\pi)$GHz and ns. $\mathcal{F}$ is the single-photon fidelity, $\mathcal{F}_{\text{en}}$ is the entangling fidelity, $\eta_{\text{abs}}$ is the probability of measuring state $s$ after photon absorption, and $\eta_d$ is the probability of correct detection via a controlled-phase gate. The cavity parameters correspond to silicon nitride microdisk optical resonators~\cite{Barclay}.
\end{minipage}
\begin{tabular}{p{2.4cm} p{0.5cm} p{0.6cm} p{0.8cm} p{1cm} p{0.8cm} p{0.8cm} p{1.2cm} p{1.2cm} p{1cm} p{1.2cm} p{1cm} p{1.1cm} p{1.6cm}}
\toprule
Cooling method & $g$ & $\kappa$ & $\kappa_i$ & $2\gamma$ & $C$ & $2\Omega_0$ & $1-\mathcal{F}$ & $1-\mathcal{F}_{\text{en}}$ & $1-\eta_{\text{abs}}$ & $1-\eta_d$ & $T$ & $\tau$ & $\tau/T$ \\
\midrule
Zeeman cooling & 1.6 & 2 & 0.01 & 0.0061 & 420 & 0.7 & 0.0098 & 0.014 & 0.0098 & 0.019 & 7.96 & 343 & 43 \\
MOT            & 1.6 & 2 & 0.01 & 0.0061 & 420 & 0.5 & 0.0098 & 0.011 & 0.0098 & 0.019 & 15.9 & 1500 & 94 \\
Dipole trap    & 1.6 & 2 & 0.01 & 0.0061 & 420 & 0.5 & 0.0098 & 0.011 & 0.0098 & 0.019 & 15.9 & $2\times 10^6$ & $1.3 \times 10^5$ \\
\midrule
\bottomrule
\end{tabular}
\label{tab3}
\vspace{0.5em}
\end{center}
\end{widetext}

 \end{document}